%% file: cryoifo.tex
\documentclass[12pt,a4paper]{iopart}
\usepackage{graphicx}
\usepackage{fouriernc}
\usepackage{iopams}
\usepackage{longtable}
\usepackage{rotating}
\usepackage{fancyhdr}
\usepackage[svgnames]{xcolor} 
\usepackage{subcaption}
\usepackage{wrapfig}
\usepackage{booktabs}
\usepackage{upgreek}
\usepackage{tablefootnote}
\usepackage{bm}                      
\usepackage{appendix}
\usepackage{url}
\usepackage{colortbl}
\usepackage{hyperref}
\hypersetup{
        colorlinks=true,
        citecolor=SteelBlue,
        filecolor=LimeGreen,
        linkcolor=SlateBlue,
        urlcolor=MediumPurple
}
\usepackage{cleveref}
\usepackage{siunitx}
\usepackage{tikz}
\usepackage{datetime}

\graphicspath{{figures/}}

\definecolor{BadRed}{rgb}{1 0.25 0.25}
\definecolor{NotSureOrng}{rgb}{1 0.647 0.5}
\definecolor{OKYellow}{rgb}{1 1 0.5}
\definecolor{GoodGreen}{rgb}{0.5 1 0.5}

\newcommand{\Tcryo}{\GwincVal{Blue.Materials.Substrate.Temp}}

\def\gw#1{gravitational wave#1 (GW#1)\gdef\gw{GW}}

\newcommand{\voy}{LIGO Voyager}
\input{BlueBird5.tex}
\newcommand{\LaserPower}{\SI[round-mode=figures,round-precision=3]{\GwincVal{Blue.Laser.Power}}{W}}
\newcommand{\LaserPowerApprox}{\SI[round-mode=figures,round-precision=2]{\GwincVal{Blue.Laser.Power}}{W}}
\newcommand{\LaserPowerPSL}{\SI[round-mode=figures,round-precision=2]{\GwincVal{Blue.Laser.PMCInput}}{W}}
\newcommand{\BSPower}{\SI[round-mode=figures,round-precision=2]{\GwincVal{Blue.Laser.BSPower_Watts}}{W}}
\newcommand{\ArmPower}{\SI[exponent-to-prefix,round-mode=figures,round-precision=1,scientific-notation=engineering]{\GwincVal{Blue.Laser.ArmPower}}{\watt}}
\newcommand{\TMDiameterCm}{\pgfmathparse{int(\GwincVal{Blue.Materials.MassRadius_cm}*2)}\pgfmathresult\,cm}

\newcommand{\SusTotalMass}{\SI[round-mode=figures,round-precision=3]{\GwincVal{Blue.Suspension.Stage(4).CummulativeMass}}{}}
\newcommand{\SusMassTop}{\SI[round-mode=figures,round-precision=3]{\GwincVal{Blue.Suspension.Stage(4).Mass}}{}}
\newcommand{\SusMassUIM}{\SI[round-mode=figures,round-precision=3]{\GwincVal{Blue.Suspension.Stage(3).Mass}}{}}
\newcommand{\SusMassPUM}{\SI[round-mode=figures,round-precision=3]{\GwincVal{Blue.Suspension.Stage(2).Mass}}{}}
\newcommand{\SusMassTM}{\SI[round-mode=figures,round-precision=3]{\GwincVal{Blue.Suspension.Stage(1).Mass}}{}}
\newcommand{\SusLengthTop}{\SI[round-mode=figures,round-precision=3]{\GwincVal{Blue.Suspension.Stage(4).Length}}{}}
\newcommand{\SusLengthUIM}{\SI[round-mode=figures,round-precision=3]{\GwincVal{Blue.Suspension.Stage(3).Length}}{}}
\newcommand{\SusLengthPUM}{\SI[round-mode=figures,round-precision=3]{\GwincVal{Blue.Suspension.Stage(2).Length}}{}}
\newcommand{\SusLengthTM}{\SI[round-mode=figures,round-precision=3]{\GwincVal{Blue.Suspension.Stage(1).Length}}{}}

\begin{document}

\title{A Cryogenic Silicon Interferometer for Gravitational-wave Detection}

\input{author.tex}
\input{affl.tex}


\begin{abstract}
The detection of gravitational waves from compact binary mergers by LIGO has
opened the era of gravitational wave astronomy, revealing a previously hidden side of the cosmos. 
To maximize the reach of the existing LIGO observatory facilities, we have designed a new instrument able to detect gravitational waves at distances 5 times further away than possible with Advanced LIGO, or at greater than 100 times the event rate. 
Observations with this new instrument will make possible dramatic steps toward understanding the physics of the nearby universe, as well as observing the universe out to cosmological distances by the detection of binary black hole coalescences. 
This article presents the instrument design and a quantitative analysis of the anticipated noise floor.


\end{abstract}

\maketitle

\tableofcontents

\clearpage
\section{Introduction}
\markboth{Introduction}{}                   
\input{intro}

\clearpage
\section{Test Masses}
\markboth{Test Masses}{}
\input{SiliconMasses}

\clearpage
\section{Optical Coatings}
\markboth{Optical Coatings}{}
\input{Coatings}

\clearpage
\section{Choice of Laser Wavelength}
\markboth{Choice of Laser Wavelength}{}
\input{Wavelength}

\clearpage
\section{Quantum Noise}
\markboth{Quantum Noise}{}
\input{Quantum}

\clearpage
\section{Suspensions}
\markboth{Suspensions}{}
\input{Suspensions}

\clearpage
\section{Laser Technology}
\markboth{Laser Technology}{}

\input{Lasers}

\clearpage
\section{Configurations}
\markboth{Configurations}{}
\input{configurations}




\input{conclusion}

\clearpage
\appendices
\appendixpage
\addappheadtotoc

\section{Cryogenics}
\input{Cryogenics}


\clearpage
\bibliographystyle{unsrt85}
\bibliography{gw_references}

\end{document}

%% file: BlueBird5.tex
\providecommand\DefGwincVal[2]{%
  \expandafter\newcommand\csname gwincval-#1\endcsname{#2}%
}
\providecommand{\GwincVal}[1]{\csname gwincval-#1\endcsname}
\DefGwincVal{Blue.Infrastructure.Length}{3995}
\DefGwincVal{Blue.Infrastructure.ResidualGas.pressure}{4e-07}
\DefGwincVal{Blue.Infrastructure.ResidualGas.mass}{3.35e-27}
\DefGwincVal{Blue.Infrastructure.ResidualGas.polarizability}{8.1e-31}
\DefGwincVal{Blue.Constants.E0}{8.8542e-12}
\DefGwincVal{Blue.Constants.hbar}{1.0546e-34}
\DefGwincVal{Blue.Constants.kB}{1.3807e-23}
\DefGwincVal{Blue.Constants.h}{6.6261e-34}
\DefGwincVal{Blue.Constants.R}{8.3145}
\DefGwincVal{Blue.Constants.m_e}{9.1094e-31}
\DefGwincVal{Blue.Constants.c}{299792458}
\DefGwincVal{Blue.Constants.Temp}{295}
\DefGwincVal{Blue.Constants.yr}{31556926.08}
\DefGwincVal{Blue.Constants.M_earth}{5.972e+24}
\DefGwincVal{Blue.Constants.R_earth}{6378100}
\DefGwincVal{Blue.Constants.fs}{16384}
\DefGwincVal{Blue.Constants.AU}{149597870700}
\DefGwincVal{Blue.Constants.parsec}{3.085677581491367e+16}
\DefGwincVal{Blue.Constants.Mpc}{3.085677581491367e+22}
\DefGwincVal{Blue.Constants.SolarMassParameter}{1.3271244e+20}
\DefGwincVal{Blue.Constants.G}{6.6741e-11}
\DefGwincVal{Blue.Constants.MSol}{1.988475415338144e+30}
\DefGwincVal{Blue.Constants.g}{9.806}
\DefGwincVal{Blue.Constants.H0}{67110}
\DefGwincVal{Blue.Constants.omegaM}{0.3175}
\DefGwincVal{Blue.Constants.omegaLambda}{0.6825}
\DefGwincVal{Blue.Constants.fInspiralMin}{3}
\DefGwincVal{Blue.Constants.BesselZeros}{[array]}
\DefGwincVal{Blue.TCS.s_cc}{7.024}
\DefGwincVal{Blue.TCS.s_cs}{7.321}
\DefGwincVal{Blue.TCS.s_ss}{7.631}
\DefGwincVal{Blue.TCS.SRCloss}{0}
\DefGwincVal{Blue.Seismic.Site}{LHO}
\DefGwincVal{Blue.Seismic.KneeFrequency}{10}
\DefGwincVal{Blue.Seismic.LowFrequencyLevel}{1e-09}
\DefGwincVal{Blue.Seismic.Gamma}{0.8}
\DefGwincVal{Blue.Seismic.Rho}{1800}
\DefGwincVal{Blue.Seismic.Beta}{0.8}
\DefGwincVal{Blue.Seismic.Omicron}{10}
\DefGwincVal{Blue.Seismic.TestMassHeight}{1.5}
\DefGwincVal{Blue.Seismic.RayleighWaveSpeed}{250}
\DefGwincVal{Blue.Seismic.darmSeiSusFile}{seismic.mat}
\DefGwincVal{Blue.Seismic.ReductionFactorPSD}{100}
\DefGwincVal{Blue.Seismic.darmseis_f}{[array]}
\DefGwincVal{Blue.Seismic.darmseis_x}{[array]}
\DefGwincVal{Blue.Suspension.BreakStress}{750000000}
\DefGwincVal{Blue.Suspension.Temp}{[array]}
\DefGwincVal{Blue.Suspension.VHCoupling.theta}{0.001}
\DefGwincVal{Blue.Suspension.Silicon.Rho}{2329}
\DefGwincVal{Blue.Suspension.Silicon.C}{300}
\DefGwincVal{Blue.Suspension.Silicon.K}{700}
\DefGwincVal{Blue.Suspension.Silicon.Alpha}{1e-10}
\DefGwincVal{Blue.Suspension.Silicon.dlnEdT}{-2e-05}
\DefGwincVal{Blue.Suspension.Silicon.Phi}{2e-09}
\DefGwincVal{Blue.Suspension.Silicon.Y}{155800000000}
\DefGwincVal{Blue.Suspension.Silicon.Dissdepth}{0.00025}
\DefGwincVal{Blue.Suspension.Silicon123K.Rho}{2329}
\DefGwincVal{Blue.Suspension.Silicon123K.C}{300}
\DefGwincVal{Blue.Suspension.Silicon123K.K}{700}
\DefGwincVal{Blue.Suspension.Silicon123K.Alpha}{1e-10}
\DefGwincVal{Blue.Suspension.Silicon123K.dlnEdT}{-2e-05}
\DefGwincVal{Blue.Suspension.Silicon123K.Phi}{2e-09}
\DefGwincVal{Blue.Suspension.Silicon123K.Y}{155800000000}
\DefGwincVal{Blue.Suspension.Silicon123K.Dissdepth}{0.00025}
\DefGwincVal{Blue.Suspension.Silicon120K.Rho}{2329}
\DefGwincVal{Blue.Suspension.Silicon120K.C}{300}
\DefGwincVal{Blue.Suspension.Silicon120K.K}{700}
\DefGwincVal{Blue.Suspension.Silicon120K.Alpha}{1e-10}
\DefGwincVal{Blue.Suspension.Silicon120K.dlnEdT}{-2e-05}
\DefGwincVal{Blue.Suspension.Silicon120K.Phi}{2e-09}
\DefGwincVal{Blue.Suspension.Silicon120K.Y}{155800000000}
\DefGwincVal{Blue.Suspension.Silicon120K.Dissdepth}{0.00025}
\DefGwincVal{Blue.Suspension.Silicon295K.Rho}{2329}
\DefGwincVal{Blue.Suspension.Silicon295K.C}{400}
\DefGwincVal{Blue.Suspension.Silicon295K.K}{148}
\DefGwincVal{Blue.Suspension.Silicon295K.Alpha}{2.6e-06}
\DefGwincVal{Blue.Suspension.Silicon295K.dlnEdT}{-7e-05}
\DefGwincVal{Blue.Suspension.Silicon295K.Y}{166000000000}
\DefGwincVal{Blue.Suspension.Silicon295K.Phi}{2e-08}
\DefGwincVal{Blue.Suspension.Silicon295K.Dissdepth}{0.00025}
\DefGwincVal{Blue.Suspension.Silicon300K.Rho}{2329}
\DefGwincVal{Blue.Suspension.Silicon300K.C}{400}
\DefGwincVal{Blue.Suspension.Silicon300K.K}{148}
\DefGwincVal{Blue.Suspension.Silicon300K.Alpha}{2.6e-06}
\DefGwincVal{Blue.Suspension.Silicon300K.dlnEdT}{-7e-05}
\DefGwincVal{Blue.Suspension.Silicon300K.Y}{166000000000}
\DefGwincVal{Blue.Suspension.Silicon300K.Phi}{2e-08}
\DefGwincVal{Blue.Suspension.Silicon300K.Dissdepth}{0.00025}
\DefGwincVal{Blue.Suspension.FiberType}{1}
\DefGwincVal{Blue.Suspension.Silica.Rho}{2200}
\DefGwincVal{Blue.Suspension.Silica.C}{772}
\DefGwincVal{Blue.Suspension.Silica.K}{1.38}
\DefGwincVal{Blue.Suspension.Silica.Alpha}{3.9e-07}
\DefGwincVal{Blue.Suspension.Silica.dlnEdT}{0.000152}
\DefGwincVal{Blue.Suspension.Silica.Phi}{4.1e-10}
\DefGwincVal{Blue.Suspension.Silica.Y}{72000000000}
\DefGwincVal{Blue.Suspension.Silica.Dissdepth}{0.015}
\DefGwincVal{Blue.Suspension.Silica300K.Rho}{2200}
\DefGwincVal{Blue.Suspension.Silica300K.C}{772}
\DefGwincVal{Blue.Suspension.Silica300K.K}{1.38}
\DefGwincVal{Blue.Suspension.Silica300K.Alpha}{3.9e-07}
\DefGwincVal{Blue.Suspension.Silica300K.dlnEdT}{0.000152}
\DefGwincVal{Blue.Suspension.Silica300K.Phi}{4.1e-10}
\DefGwincVal{Blue.Suspension.Silica300K.Y}{72000000000}
\DefGwincVal{Blue.Suspension.Silica300K.Dissdepth}{0.015}
\DefGwincVal{Blue.Suspension.Silica295K.Rho}{2200}
\DefGwincVal{Blue.Suspension.Silica295K.C}{772}
\DefGwincVal{Blue.Suspension.Silica295K.K}{1.38}
\DefGwincVal{Blue.Suspension.Silica295K.Alpha}{3.9e-07}
\DefGwincVal{Blue.Suspension.Silica295K.dlnEdT}{0.000152}
\DefGwincVal{Blue.Suspension.Silica295K.Phi}{4.1e-10}
\DefGwincVal{Blue.Suspension.Silica295K.Y}{72000000000}
\DefGwincVal{Blue.Suspension.Silica295K.Dissdepth}{0.015}
\DefGwincVal{Blue.Suspension.Silica123K.Rho}{2200}
\DefGwincVal{Blue.Suspension.Silica123K.C}{300}
\DefGwincVal{Blue.Suspension.Silica123K.K}{0.75}
\DefGwincVal{Blue.Suspension.Silica123K.Alpha}{-5e-07}
\DefGwincVal{Blue.Suspension.Silica123K.dlnEdT}{0.000152}
\DefGwincVal{Blue.Suspension.Silica123K.Y}{70000000000}
\DefGwincVal{Blue.Suspension.Silica123K.Phi}{2e-05}
\DefGwincVal{Blue.Suspension.Silica123K.Dissdepth}{0.015}
\DefGwincVal{Blue.Suspension.C70Steel.Rho}{7800}
\DefGwincVal{Blue.Suspension.C70Steel.C}{486}
\DefGwincVal{Blue.Suspension.C70Steel.K}{49}
\DefGwincVal{Blue.Suspension.C70Steel.Alpha}{1.2e-05}
\DefGwincVal{Blue.Suspension.C70Steel.dlnEdT}{-0.00025}
\DefGwincVal{Blue.Suspension.C70Steel.Phi}{0.0002}
\DefGwincVal{Blue.Suspension.C70Steel.Y}{212000000000}
\DefGwincVal{Blue.Suspension.C70Steel295K.Rho}{7800}
\DefGwincVal{Blue.Suspension.C70Steel295K.C}{486}
\DefGwincVal{Blue.Suspension.C70Steel295K.K}{49}
\DefGwincVal{Blue.Suspension.C70Steel295K.Alpha}{1.2e-05}
\DefGwincVal{Blue.Suspension.C70Steel295K.dlnEdT}{-0.00025}
\DefGwincVal{Blue.Suspension.C70Steel295K.Phi}{0.0002}
\DefGwincVal{Blue.Suspension.C70Steel295K.Y}{212000000000}
\DefGwincVal{Blue.Suspension.C70Steel300K.Rho}{7800}
\DefGwincVal{Blue.Suspension.C70Steel300K.C}{486}
\DefGwincVal{Blue.Suspension.C70Steel300K.K}{49}
\DefGwincVal{Blue.Suspension.C70Steel300K.Alpha}{1.2e-05}
\DefGwincVal{Blue.Suspension.C70Steel300K.dlnEdT}{-0.00025}
\DefGwincVal{Blue.Suspension.C70Steel300K.Phi}{0.0002}
\DefGwincVal{Blue.Suspension.C70Steel300K.Y}{212000000000}
\DefGwincVal{Blue.Suspension.C70Steel123K.Rho}{7800}
\DefGwincVal{Blue.Suspension.C70Steel123K.C}{250}
\DefGwincVal{Blue.Suspension.C70Steel123K.K}{15}
\DefGwincVal{Blue.Suspension.C70Steel123K.Alpha}{8e-06}
\DefGwincVal{Blue.Suspension.C70Steel123K.dlnEdT}{-0.00025}
\DefGwincVal{Blue.Suspension.C70Steel123K.Phi}{0.0002}
\DefGwincVal{Blue.Suspension.C70Steel123K.Y}{212000000000}
\DefGwincVal{Blue.Suspension.MaragingSteel.Rho}{7800}
\DefGwincVal{Blue.Suspension.MaragingSteel.C}{460}
\DefGwincVal{Blue.Suspension.MaragingSteel.K}{20}
\DefGwincVal{Blue.Suspension.MaragingSteel.Alpha}{1.1e-05}
\DefGwincVal{Blue.Suspension.MaragingSteel.dlnEdT}{0}
\DefGwincVal{Blue.Suspension.MaragingSteel.Phi}{0.0001}
\DefGwincVal{Blue.Suspension.MaragingSteel.Y}{187000000000}
\DefGwincVal{Blue.Suspension.MaragingSteel295K.Rho}{7800}
\DefGwincVal{Blue.Suspension.MaragingSteel295K.C}{460}
\DefGwincVal{Blue.Suspension.MaragingSteel295K.K}{20}
\DefGwincVal{Blue.Suspension.MaragingSteel295K.Alpha}{1.1e-05}
\DefGwincVal{Blue.Suspension.MaragingSteel295K.dlnEdT}{0}
\DefGwincVal{Blue.Suspension.MaragingSteel295K.Phi}{0.0001}
\DefGwincVal{Blue.Suspension.MaragingSteel295K.Y}{187000000000}
\DefGwincVal{Blue.Suspension.MaragingSteel300K.Rho}{7800}
\DefGwincVal{Blue.Suspension.MaragingSteel300K.C}{460}
\DefGwincVal{Blue.Suspension.MaragingSteel300K.K}{20}
\DefGwincVal{Blue.Suspension.MaragingSteel300K.Alpha}{1.1e-05}
\DefGwincVal{Blue.Suspension.MaragingSteel300K.dlnEdT}{0}
\DefGwincVal{Blue.Suspension.MaragingSteel300K.Phi}{0.0001}
\DefGwincVal{Blue.Suspension.MaragingSteel300K.Y}{187000000000}
\DefGwincVal{Blue.Suspension.Type}{BQuad}
\DefGwincVal{Blue.Suspension.Stage(1).Length}{0.78235}
\DefGwincVal{Blue.Suspension.Stage(1).Mass}{200}
\DefGwincVal{Blue.Suspension.Stage(1).CummulativeMass}{200}
\DefGwincVal{Blue.Suspension.Stage(1).Dilution}{NaN}
\DefGwincVal{Blue.Suspension.Stage(1).K}{260000}
\DefGwincVal{Blue.Suspension.Stage(1).NWires}{4}
\DefGwincVal{Blue.Suspension.Stage(1).WireRadius}{NaN}
\DefGwincVal{Blue.Suspension.Stage(1).Blade}{0.012}
\DefGwincVal{Blue.Suspension.Stage(1).WireMaterial}{Silicon}
\DefGwincVal{Blue.Suspension.Stage(1).BladeMaterial}{Silicon}
\DefGwincVal{Blue.Suspension.Stage(2).Length}{0.5592}
\DefGwincVal{Blue.Suspension.Stage(2).Mass}{200}
\DefGwincVal{Blue.Suspension.Stage(2).CummulativeMass}{400}
\DefGwincVal{Blue.Suspension.Stage(2).Dilution}{NaN}
\DefGwincVal{Blue.Suspension.Stage(2).K}{26262.6263}
\DefGwincVal{Blue.Suspension.Stage(2).NWires}{4}
\DefGwincVal{Blue.Suspension.Stage(2).WireRadius}{0.00066776}
\DefGwincVal{Blue.Suspension.Stage(2).Blade}{0.007206}
\DefGwincVal{Blue.Suspension.Stage(2).WireMaterial}{C70Steel}
\DefGwincVal{Blue.Suspension.Stage(2).BladeMaterial}{MaragingSteel}
\DefGwincVal{Blue.Suspension.Stage(3).Length}{0.15}
\DefGwincVal{Blue.Suspension.Stage(3).Mass}{70.0417}
\DefGwincVal{Blue.Suspension.Stage(3).CummulativeMass}{470.0417}
\DefGwincVal{Blue.Suspension.Stage(3).Dilution}{NaN}
\DefGwincVal{Blue.Suspension.Stage(3).K}{18150.1268}
\DefGwincVal{Blue.Suspension.Stage(3).NWires}{4}
\DefGwincVal{Blue.Suspension.Stage(3).WireRadius}{0.00072387}
\DefGwincVal{Blue.Suspension.Stage(3).Blade}{0.00768}
\DefGwincVal{Blue.Suspension.Stage(3).WireMaterial}{C70Steel}
\DefGwincVal{Blue.Suspension.Stage(3).BladeMaterial}{MaragingSteel}
\DefGwincVal{Blue.Suspension.Stage(4).Length}{0.15045}
\DefGwincVal{Blue.Suspension.Stage(4).Mass}{49.9583}
\DefGwincVal{Blue.Suspension.Stage(4).CummulativeMass}{520}
\DefGwincVal{Blue.Suspension.Stage(4).Dilution}{NaN}
\DefGwincVal{Blue.Suspension.Stage(4).K}{114362.3071}
\DefGwincVal{Blue.Suspension.Stage(4).NWires}{2}
\DefGwincVal{Blue.Suspension.Stage(4).WireRadius}{0.0010767}
\DefGwincVal{Blue.Suspension.Stage(4).Blade}{0.01388}
\DefGwincVal{Blue.Suspension.Stage(4).WireMaterial}{C70Steel}
\DefGwincVal{Blue.Suspension.Stage(4).BladeMaterial}{MaragingSteel}
\DefGwincVal{Blue.Suspension.Ribbon.Thickness}{0.0005}
\DefGwincVal{Blue.Suspension.Ribbon.Width}{0.01}
\DefGwincVal{Blue.Suspension.hForce}{[array]}
\DefGwincVal{Blue.Suspension.vForce}{[array]}
\DefGwincVal{Blue.Suspension.hForce_singlylossy}{[array]}
\DefGwincVal{Blue.Suspension.vForce_singlylossy}{[array]}
\DefGwincVal{Blue.Suspension.hTable}{[array]}
\DefGwincVal{Blue.Suspension.vTable}{[array]}
\DefGwincVal{Blue.Materials.Coating.Yhighn}{60000000000}
\DefGwincVal{Blue.Materials.Coating.Sigmahighn}{0.22}
\DefGwincVal{Blue.Materials.Coating.CVhighn}{1050000}
\DefGwincVal{Blue.Materials.Coating.Alphahighn}{1e-09}
\DefGwincVal{Blue.Materials.Coating.Betahighn}{0.00014}
\DefGwincVal{Blue.Materials.Coating.ThermalDiffusivityhighn}{1.03}
\DefGwincVal{Blue.Materials.Coating.Phihighn}{1e-05}
\DefGwincVal{Blue.Materials.Coating.Indexhighn}{3.5}
\DefGwincVal{Blue.Materials.Coating.Ylown}{72000000000}
\DefGwincVal{Blue.Materials.Coating.Sigmalown}{0.17}
\DefGwincVal{Blue.Materials.Coating.CVlown}{1641200}
\DefGwincVal{Blue.Materials.Coating.Alphalown}{5.1e-07}
\DefGwincVal{Blue.Materials.Coating.Betalown}{8e-06}
\DefGwincVal{Blue.Materials.Coating.ThermalDiffusivitylown}{1.05}
\DefGwincVal{Blue.Materials.Coating.Philown}{0.0001}
\DefGwincVal{Blue.Materials.Coating.Indexlown}{1.436}
\DefGwincVal{Blue.Materials.Substrate.c2}{3e-13}
\DefGwincVal{Blue.Materials.Substrate.MechanicalLossExponent}{1}
\DefGwincVal{Blue.Materials.Substrate.Alphas}{5.2e-12}
\DefGwincVal{Blue.Materials.Substrate.MirrorY}{155800000000}
\DefGwincVal{Blue.Materials.Substrate.MirrorSigma}{0.27}
\DefGwincVal{Blue.Materials.Substrate.MassDensity}{2329}
\DefGwincVal{Blue.Materials.Substrate.MassAlpha}{1e-09}
\DefGwincVal{Blue.Materials.Substrate.MassCM}{300}
\DefGwincVal{Blue.Materials.Substrate.MassKappa}{700}
\DefGwincVal{Blue.Materials.Substrate.RefractiveIndex}{3.5}
\DefGwincVal{Blue.Materials.Substrate.dndT}{0.0001}
\DefGwincVal{Blue.Materials.Substrate.Temp}{123}
\DefGwincVal{Blue.Materials.Substrate.isSemiConductor}{1}
\DefGwincVal{Blue.Materials.Substrate.CarrierDensity}{1e+19}
\DefGwincVal{Blue.Materials.Substrate.ElectronDiffusion}{0.0097}
\DefGwincVal{Blue.Materials.Substrate.HoleDiffusion}{0.0035}
\DefGwincVal{Blue.Materials.Substrate.ElectronEffMass}{9.747e-31}
\DefGwincVal{Blue.Materials.Substrate.HoleEffMass}{8.0163e-31}
\DefGwincVal{Blue.Materials.Substrate.ElectronIndexGamma}{-8.8e-28}
\DefGwincVal{Blue.Materials.Substrate.HoleIndexGamma}{-1.02e-27}
\DefGwincVal{Blue.Materials.MassRadius}{0.225}
\DefGwincVal{Blue.Materials.MassThickness}{0.55}
\DefGwincVal{Blue.Materials.MirrorVolume}{0.087474}
\DefGwincVal{Blue.Materials.MirrorMass}{203.7263}
\DefGwincVal{Blue.Laser.Wavelength}{2e-06}
\DefGwincVal{Blue.Laser.Power}{151.5919}
\DefGwincVal{Blue.Laser.ArmPower}{3050535.1816}
\DefGwincVal{Blue.Optics.Type}{SignalRecycled}
\DefGwincVal{Blue.Optics.SRM.CavityLength}{55}
\DefGwincVal{Blue.Optics.SRM.Transmittance}{0.045773}
\DefGwincVal{Blue.Optics.SRM.Tunephase}{0}
\DefGwincVal{Blue.Optics.PhotoDetectorEfficiency}{0.95}
\DefGwincVal{Blue.Optics.Loss}{1e-05}
\DefGwincVal{Blue.Optics.BSLoss}{0.0005}
\DefGwincVal{Blue.Optics.coupling}{1}
\DefGwincVal{Blue.Optics.Curvature.ITM}{1800}
\DefGwincVal{Blue.Optics.Curvature.ETM}{2500}
\DefGwincVal{Blue.Optics.SubstrateAbsorption}{3e-07}
\DefGwincVal{Blue.Optics.ITM.SubstrateAbsorption}{0.001}
\DefGwincVal{Blue.Optics.ITM.BeamRadius}{0.058504}
\DefGwincVal{Blue.Optics.ITM.CoatingAbsorption}{1e-06}
\DefGwincVal{Blue.Optics.ITM.Transmittance}{0.002}
\DefGwincVal{Blue.Optics.ITM.CoatingThicknessLown}{0.308}
\DefGwincVal{Blue.Optics.ITM.CoatingThicknessCap}{0.5}
\DefGwincVal{Blue.Optics.ITM.CoatLayerOpticalThickness}{[array]}
\DefGwincVal{Blue.Optics.ITM.Thickness}{0.55}
\DefGwincVal{Blue.Optics.ITM.Pabs}{[array]}
\DefGwincVal{Blue.Optics.pcrit}{10}
\DefGwincVal{Blue.Optics.ETM.BeamRadius}{0.083543}
\DefGwincVal{Blue.Optics.ETM.Transmittance}{5e-06}
\DefGwincVal{Blue.Optics.ETM.CoatingThicknessLown}{0.27}
\DefGwincVal{Blue.Optics.ETM.CoatingThicknessCap}{0.5}
\DefGwincVal{Blue.Optics.ETM.CoatLayerOpticalThickness}{[array]}
\DefGwincVal{Blue.Optics.PRM.Transmittance}{0.049202}
\DefGwincVal{Blue.Optics.Quadrature.dc}{1.5832}
\DefGwincVal{Blue.Squeezer.Type}{Freq Dependent}
\DefGwincVal{Blue.Squeezer.AmplitudedB}{10}
\DefGwincVal{Blue.Squeezer.InjectionLoss}{0.05}
\DefGwincVal{Blue.Squeezer.SQZAngle}{0}
\DefGwincVal{Blue.Squeezer.FilterCavity.fdetune}{-25}
\DefGwincVal{Blue.Squeezer.FilterCavity.L}{300}
\DefGwincVal{Blue.Squeezer.FilterCavity.Ti}{0.0006149}
\DefGwincVal{Blue.Squeezer.FilterCavity.Te}{0}
\DefGwincVal{Blue.Squeezer.FilterCavity.Lrt}{1e-05}
\DefGwincVal{Blue.Squeezer.FilterCavity.Rot}{0}
\DefGwincVal{Blue.OutputFilter.Type}{None}
\DefGwincVal{Blue.OutputFilter.FilterCavity.fdetune}{-30}
\DefGwincVal{Blue.OutputFilter.FilterCavity.L}{4000}
\DefGwincVal{Blue.OutputFilter.FilterCavity.Ti}{0.01}
\DefGwincVal{Blue.OutputFilter.FilterCavity.Te}{0}
\DefGwincVal{Blue.OutputFilter.FilterCavity.Lrt}{0.0001}
\DefGwincVal{Blue.OutputFilter.FilterCavity.Rot}{0}
\DefGwincVal{Blue.Model}{IFOModel_123_2000}
\DefGwincVal{Blue.modeSR}{0}
\DefGwincVal{Blue.gwinc.PRfixed}{0}
\DefGwincVal{Blue.gwinc.pbs}{3081.0405}
\DefGwincVal{Blue.gwinc.parm}{3002539.9267}
\DefGwincVal{Blue.gwinc.finesse}{3110.4878}
\DefGwincVal{Blue.gwinc.prfactor}{20.3246}
\DefGwincVal{Blue.gwinc.fSQL}{38.2924}
\DefGwincVal{Blue.gwinc.fGammaIFO}{509.8349}
\DefGwincVal{Blue.gwinc.fGammaArm}{5.9716}
\DefGwincVal{Blue.nse.ResGas}{[array]}
\DefGwincVal{Blue.nse.SuspThermal}{[array]}
\DefGwincVal{Blue.nse.Quantum}{[array]}
\DefGwincVal{Blue.nse.Freq}{[array]}
\DefGwincVal{Blue.nse.Newtonian}{[array]}
\DefGwincVal{Blue.nse.Seismic}{[array]}
\DefGwincVal{Blue.nse.Total}{[array]}
\DefGwincVal{Blue.nse.MirrorThermal.Total}{[array]}
\DefGwincVal{Blue.nse.MirrorThermal.SubBrown}{[array]}
\DefGwincVal{Blue.nse.MirrorThermal.CoatBrown}{[array]}
\DefGwincVal{Blue.nse.MirrorThermal.SubTE}{[array]}
\DefGwincVal{Blue.nse.MirrorThermal.CoatTO}{[array]}
\DefGwincVal{Blue.nse.MirrorThermal.SubTR}{[array]}
\DefGwincVal{Blue.nse.MirrorThermal.SubCD}{[array]}
\DefGwincVal{Blue.Laser.Wavelength_nm}{2000}
\DefGwincVal{Blue.Laser.ArmPower_Watts}{3050535}
\DefGwincVal{Blue.Laser.BSPower_Watts}{3081}
\DefGwincVal{Blue.Laser.IMCOutput}{152}
\DefGwincVal{Blue.Laser.IMCInput}{202}
\DefGwincVal{Blue.Laser.PMCOutput}{212}
\DefGwincVal{Blue.Laser.PMCInput}{222}
\DefGwincVal{Blue.Laser.ITMIntensity}{89127}
\DefGwincVal{Blue.Laser.BSIntensity}{90}
\DefGwincVal{Blue.Materials.MassRadius_cm}{22.5}
\DefGwincVal{Blue.Materials.MassThickness_cm}{55}
\DefGwincVal{Blue.Optics.ITM.BeamRadius_cm}{5.9}
\DefGwincVal{Blue.Optics.ETM.BeamRadius_cm}{8.4}
\DefGwincVal{Blue.Substance}{silicon}
\DefGwincVal{Blue.Suspension.FiberType_str}{ribbon}
\DefGwincVal{Blue.Squeezer.FilterCavity.Lrt_ppm}{10}

%% file: author.tex
\author{%
R.~X.~Adhikari$^{1,*}$, 
K.~Arai$^{1}$, 
A.~F.~Brooks$^{1}$, 
C.~Wipf$^{1}$, 
O.~Aguiar$^{2}$, 
P.~Altin$^{3}$, 
B.~Barr$^{4}$, 
L.~Barsotti$^{5}$, 
R.~Bassiri$^{6}$, 
A.~Bell$^{4}$
% ...
}

%% file: affl.tex
\address {$^{1}$LIGO, California Institute of Technology, Pasadena, CA 91125, USA } 
\address {$^{2}$Instituto Nacional de Pesquisas Espaciais, 12227-010 S\~{a}o Jos\'{e} dos Campos, S\~{a}o Paulo, Brazil } 
\address {$^{3}$OzGrav, ANU Centre for Gravitational Astrophysics, Research Schools of Physics, and Astronomy and Astrophysics, The Australian National University, Canberra, 2601, Australia } 
\address {$^{4}$SUPA, University of Glasgow, Glasgow G12 8QQ, UK } 
\address {$^{5}$LIGO, Massachusetts Institute of Technology, Cambridge, MA 02139, USA } 
\address {$^{6}$Stanford University, Stanford, CA 94305, USA } 
\address {$^{7}$SUPA, University of the West of Scotland, Paisley Scotland PA1 2BE, UK } 
\address {$^{8}$OzGrav, University of Western Australia, Crawley, Western Australia 6009, Australia } 
\address {$^{9}$OzGrav, University of Adelaide, Adelaide, South Australia 5005, Australia } 
\address {$^{10}$University of Birmingham, Birmingham B15 2TT, UK } 
\address {$^{11}$Louisiana State University, Baton Rouge, LA 70803, USA } 
\address {$^{12}$The University of Sheffield, Sheffield S10 2TN, UK } 
\address {$^{13}$LIGO Livingston Observatory, Livingston, LA 70754, USA } 
\address {$^{14}$University of Florida, Gainesville, FL 32611, USA } 
\address {$^{15}$Cardiff University, Cardiff CF24 3AA, UK } 
\address {$^{16}$Gran Sasso Science Institute (GSSI), I-67100 L'Aquila, Italy } 
\address {$^{17}$American University, Washington, D.C. 20016, USA } 
\address {$^{18}$Institut f\"{u}r Festk\"{o}rperphysik, Friedrich-Schiller-Universit\"{a}t Jena, D-07743 Jena, Germany } 
\address {$^{19}$University of California, Berkeley, CA 94720, USA } 
\address {$^{20}$Max Planck Institute for Gravitational Physics (Albert Einstein Institute), D-30167 Hannover, Germany } 
\address {$^{21}$Maastricht University, Duboisdomein 30, Maastrich Limburg 6200MD, Netherlands } 
\address {$^{22}$Faculty of Physics, Lomonosov Moscow State University, Moscow 119991, Russia } 
\address {$^{23}$Universit\"{a}t Hamburg, D-22761 Hamburg, Germany } 
\address {$^{24}$Montclair State University, Montclair, NJ 07043, USA } 
\address {$^{25}$The University of Texas Rio Grande Valley, Brownsville, TX 78520, USA } 
\address {$^{26}$SUPA, University of Strathclyde, Glasgow G1 1XQ, United Kingdom } 
\address {$^{27}$OzGrav, School of Physics and Astronomy, Monash University, Clayton 3800, Victoria, Australia } 
\address {$^{28}$National Institute of Standards and Technology (NIST), 100 Bureau Drive Stop 8171, Gaithersburg, MD 20899, USA } 
\address {$^{29}$California State University Fullerton, Fullerton, CA 92831, USA } 
\address {$^*$ Corresponding author: {\it rana@caltech.edu} }

%% file: intro.tex
The first detection of gravitational waves (GW) from the object
GW150914~\cite{GW150914} by the Advanced LIGO (aLIGO) detectors
inaugurated a new field of study: gravitational wave astronomy. The
subsequent detection of a binary neutron star
merger~\cite{MMA_BNS:2017} has highlighted the possibilities of this
new field.

GW detectors provide a probe of physics in a new regime. They offer
the best information about the extremely warped spacetime around black
holes, exotic nuclear matter in neutron stars, and, within the next
decade, a unique probe of cosmology at high redshifts.

The current LIGO detectors will approach the thermodynamic and quantum
mechanical limits of their designs within a few years. Over the next
several years, aLIGO will undergo a modest upgrade, designated
``A+''. The aim of this upgrade is chiefly to lower the quantum (shot)
noise through the use of squeezed light, and also to reduce somewhat
the thermal noise from the mirror coatings. This upgrade has the goal of enhancing the sensitivity by
$\sim$50\% \cite{DoublingRange:2015}.

In this article, we describe a more substantial upgrade, called
``\voy{}'', that will increase the range by a factor of 4\,--\,5 over
aLIGO, and the event rate by approximately 100 times, to roughly one
detection per hour. Such a dramatic change in the sensitivity should
increase the detection rate of binary neutron star mergers to about
10 per day and the rate of binary black hole mergers to around 30 per day.
This upgraded instrument would be able to detect binary black holes
out to a redshift of 8.

The path to \voy{} requires reducing several noise sources, including:
\begin{enumerate}
\item quantum radiation pressure and shot noise,
\item mirror thermal noise,
\item mirror suspension thermal noise,
\item Newtonian gravity noise
\end{enumerate}
All of these noise sources are addressed by the \voy{} design, with the goal of commissioning and observational runs within a decade.

\subsection{Justification}
The most significant design changes in \voy{} versus Advanced LIGO can be traced to the need to reduce the quantum noise in tandem with the mirror thermal noise.

\begin{itemize}
\item Quantum noise will be reduced by increasing the optical power stored in the arms.  In Advanced LIGO, the stored power is limited by thermally induced wavefront distortion effects in the fused silica test masses.  These effects will be alleviated by choosing a test mass material with a high thermal conductivity, such as silicon.

\item The test mass temperature will be lowered to \Tcryo{}\,K, to mitigate thermo-elastic noise.  This species of thermal noise is especially problematic in test masses that are good thermal conductors.  Fortunately, in silicon at \Tcryo{}\,K, the thermal expansion coefficient crosses zero, which eliminates thermo-elastic noise.  (Other plausible material candidates, such as sapphire, require cooling to near 0\,K to be free of this noise.)

\item The thermal noise of the mirror coating will be reduced by switching to low dissipation amorphous silicon based coatings, and by reducing the temperature.  Achieving low optical absorption in the amorphous silicon coatings requires an increased laser wavelength.
\end{itemize}

\begin{figure}
  \centering
  \includegraphics[width=\columnwidth]{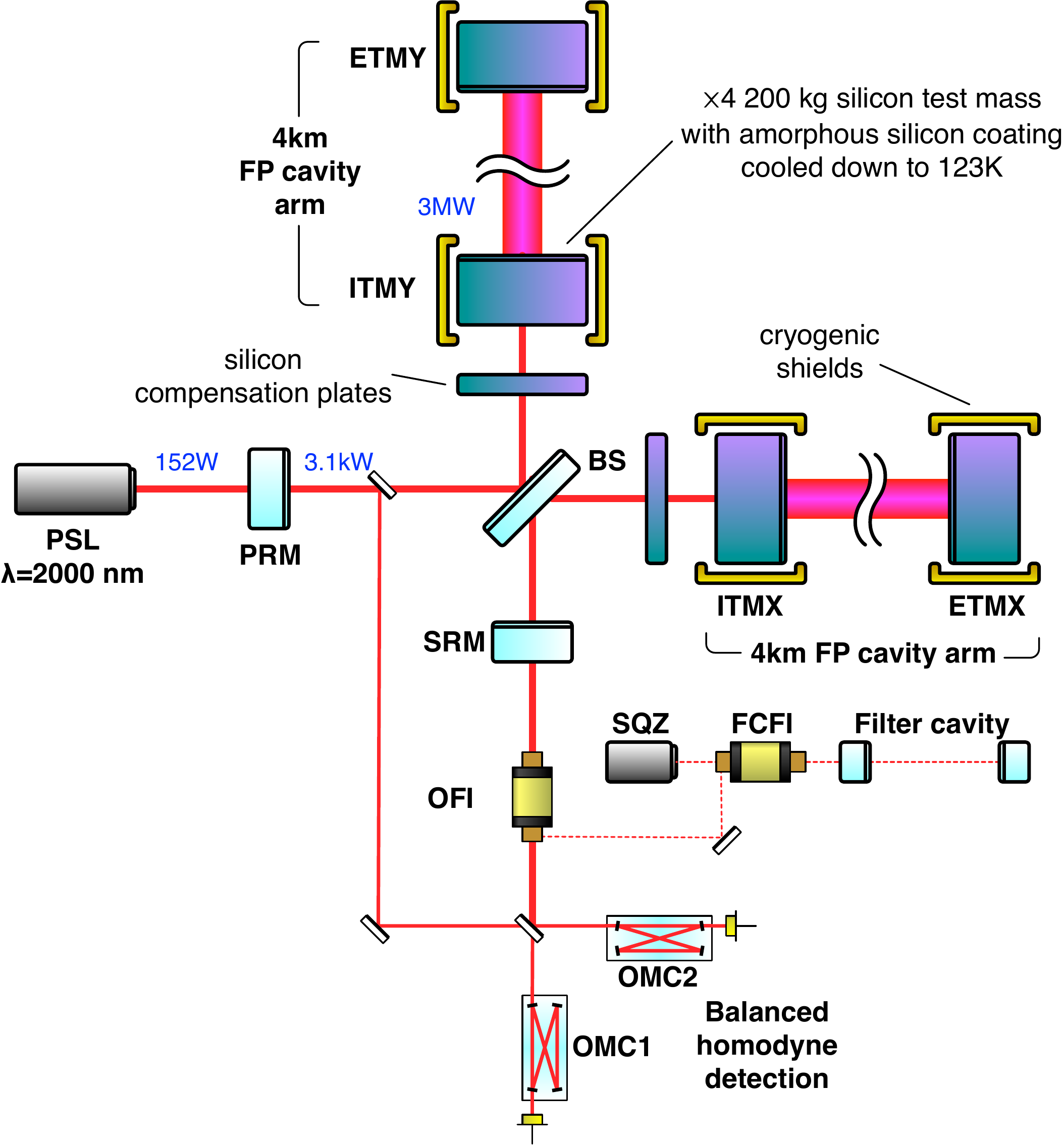}
  \caption[Schematic layout of Voyager]{A simplified schematic layout of \voy{}. Dual-recycled Fabry-Perot Michelson (DRFPMI) with frequency dependent squeezed light injection. The beam from a 2\si{\micro\meter} pre-stabilized laser (PSL), passes through an input mode cleaner (IMC) and is injected into the DRFPMI via the power-recycling mirror (PRM). Signal bandwidth is shaped via the signal recycling mirror (SRM). A squeezed vacuum source (SQZ) injects this vacuum into the DRFPMI via an output Faraday isolator (OFI) after it is reflected off a filter-cavity to provide frequency dependent squeezing. A Faraday isolator (FCFI) facilitates this coupling to the filter cavity. The output from the DRFPMI is incident on a balanced homodyne detector, which employs two output mode cleaner cavities (OMC1 and OMC2) and the local oscillator light picked off from the DRFPMI. Cold shields surround the input and end test masses in both the X and Y arms (ITMX, ITMY, ETMX and ETMY) to maintain a temperature of \Tcryo{}\,K in these optics. The high-reflectivity coatings of the test masses are made from amorphous silicon. }
  \label{fig:IFO_schematic}
\end{figure}

\begin{figure}[t]
  \centering
  \includegraphics[width=\columnwidth]{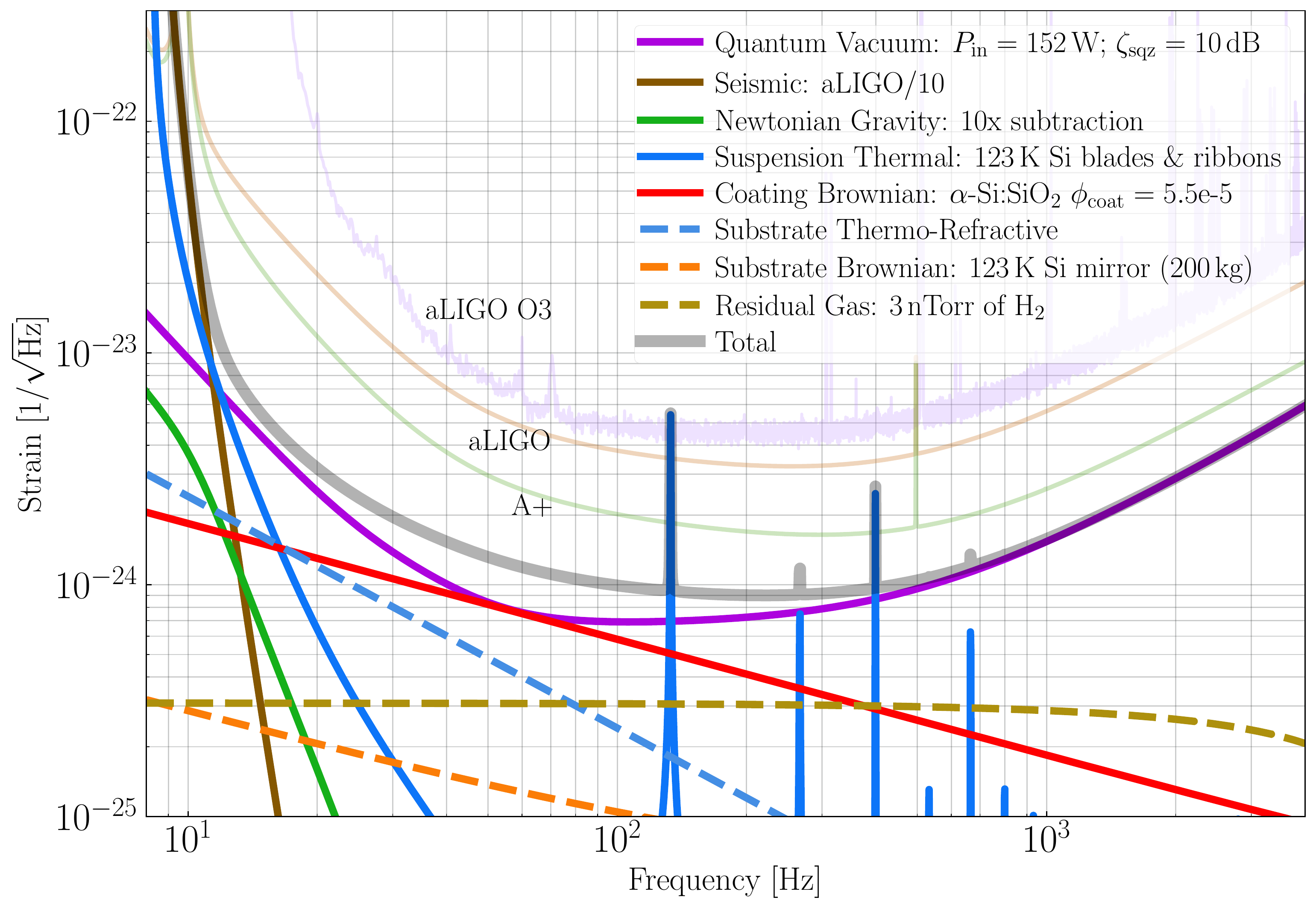}
  \caption[Voyager Noise Curve w/ Comparisons]
  {\voy{} noise curve compared to Advanced LIGO during O3,
  and the Advanced LIGO and A+ design goals.}
  \label{fig:noise_comparison}
\end{figure}

\subsection{Design overview}
The \voy{} design is illustrated in \Cref{fig:IFO_schematic}, with critical parameters called out in \Cref{tab:params}. The dual-recycled, Fabry-Perot Michelson topology is similar to Advanced LIGO and A+, with the following additional upgrades.
Optical coatings on the cryogenically-cooled (\Tcryo{}\,K) test masses will be made from amorphous silicon, with the lower coating mechanical loss and cryogenic operation reducing the coating thermal noise.
The 200\,kg test-masses will be made of crystalline silicon (rather than fused silica). The absorption spectrum of the test mass materials requires us to choose a longer wavelength laser.
The longer wavelength will also significantly reduce optical scattering from the mirrors, lowering losses and allowing for higher finesse arm cavities.
The quantum noise (shot noise and radiation pressure) will be reduced by a combination of frequency-dependent squeezing, heavier test masses, and higher stored power in the arms.
Finally, the environmentally produced Newtonian gravitational noise~\cite{Harms2015} will be reduced using seismometer arrays combined with adaptive noise regression~\cite{Cella2000, PhysRevD.86.102001}.

\begin{figure}
\centering
\begin{subfigure}[b]{0.65\textwidth}
   \includegraphics[width=1\linewidth]{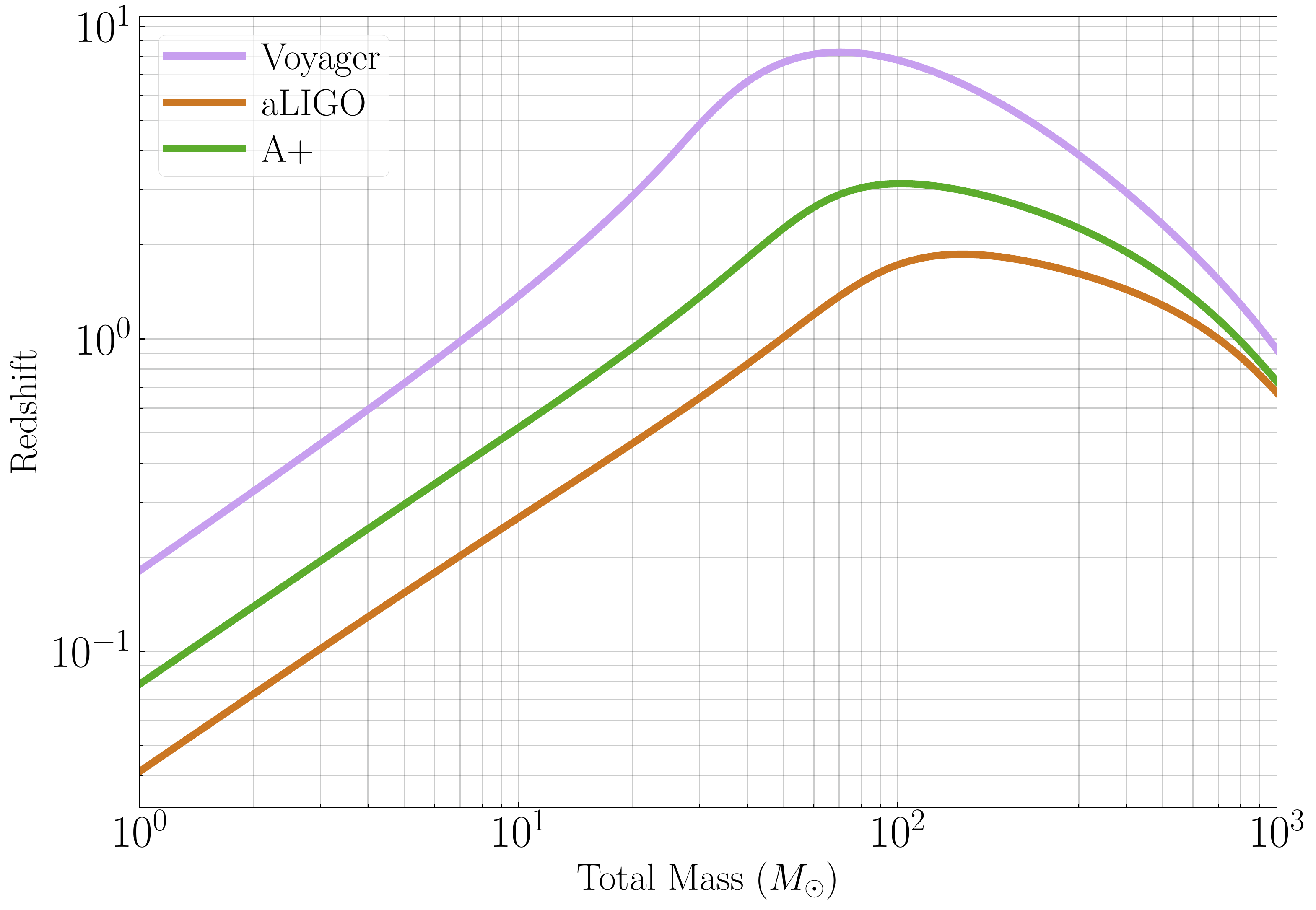}
   \caption{}
   \label{fig:RedshiftRange}
\end{subfigure}

\begin{subfigure}[b]{0.65\textwidth}
   \includegraphics[width=1\linewidth]{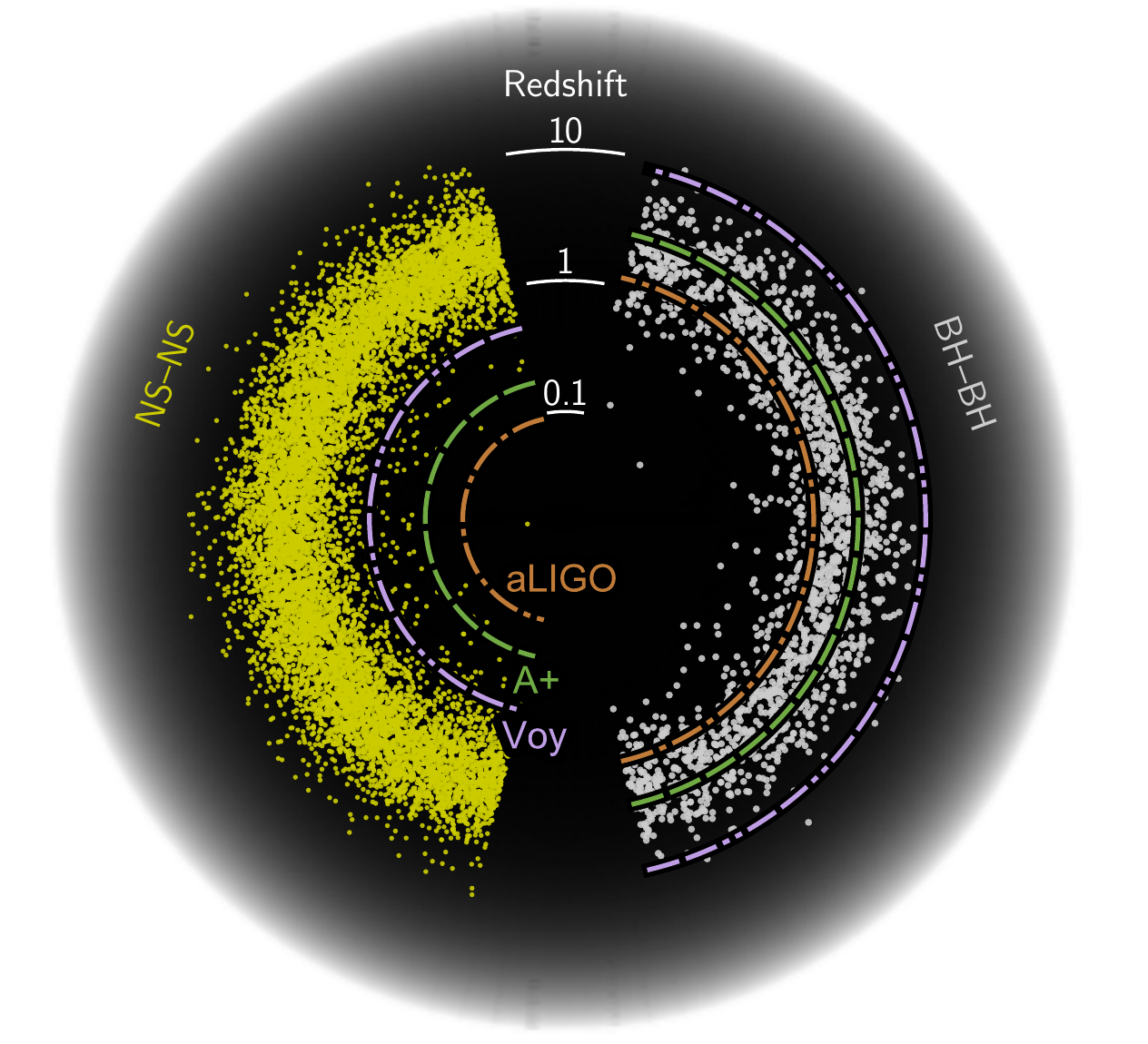}
   \caption{}
   \label{fig:horizon_donut}
\end{subfigure}
  \caption[Redshift Range]{(A) Distance at which an optimally oriented, equal mass, binary black hole merger can be detected (with SNR = 8) as a function of the total mass of the binary (in the source frame). (B) Donut visualization of the horizon distance of \voy{}, aLIGO, and A+, shown with a population of binary neutron star mergers (yellow) and 30--30 $M_\odot$ binary black hole mergers (gray). This assumes a Madau-Dickinson star formation rate~\cite{MDSFR} and a typical merger time of 100 Myr.}

\end{figure}

The \voy{} noise budget and resulting design sensitivity are shown in \Cref{fig:noise_comparison}. Horizon distances for astrophysical sources are illustrated in \Cref{fig:RedshiftRange} and \Cref{fig:horizon_donut}, showing the improvement over the Advanced LIGO design.

Although most optical components will need to be changed to handle the new wavelength, we plan on reusing the Advanced LIGO hardware and infrastructure wherever possible (for example, the seismic isolation platforms, vacuum systems, electronics and infrastructure).

\input{paramtable}

\subsection{Article overview}
This article presents a detailed description of the \voy{} design with the goals of (a) investigating the feasibility of all the required technology, largely illustrated in \Cref{fig:IFO_schematic}, and highlighting those technological areas that require further research and (b) describing all the  key noise contributions illustrated in the noise budget in \Cref{fig:noise_comparison} (and thus determining the \voy{} sensitivity).

The structure of the paper is as follows.
In \Cref{s:SiliconMasses}, we examine the feasibility of using large, cryogenically-cooled (\Tcryo{}\,K) silicon test masses and identify the substrate thermo-refractive noise, shown in the noise budget, as the limiting noise source associated with the test mass.
\Cref{s:Coatings} describes an amorphous-silicon based coating design that delivers the coating Brownian noise curve shown in the noise budget and also identifies coating absorption as a key obstacle that must be overcome.
The numerous factors that enter into the choice of \SI{2000}{\nano\meter} as the laser wavelength are described in detail in \Cref{s:Wavelength}.
Quantum noise as a limiting noise source and the feasibility of injecting \GwincVal{Blue.Squeezer.AmplitudedB}\,dB of frequency-dependent squeezed vacuum at \GwincVal{Blue.Laser.Wavelength_nm}\,nm are considered in \Cref{s:Quantum}.
The suspension thermal noise (associated with the use of silicon blades and ribbons) is described in \Cref{s:Suspensions}. This section also explores the practicality of manufacturing these silicon blades and ribbons.
In \Cref{s:Lasers}, we review the development of mid-IR laser sources and find no significant impediment to producing a thulium- or holmium-based \LaserPowerPSL{}, low-noise, single-frequency,  \SI{2000}{\nano\meter} laser within the next 10 years.
\Cref{s:configs} explores configurations of \voy{} that are optimized for high-frequency astrophysical sources, given the considerable tunability of the quantum noise curve and interferometer optical configuration.
Finally, cryogenic considerations are discussed in \Cref{s:Cryo}.

%% file: paramtable.tex

\begin{table}[tbhp]
  \centering
  {%
  \begin{tabular}{lc}
    \toprule

    Parameter & Nominal value \\

    \midrule

    Laser wavelength &
    \SI{\GwincVal{Blue.Laser.Wavelength_nm}}{nm} \\

    Laser power incident on PRM & \LaserPower \\
    
    Power in PRC & \BSPower \\
    
    Arm power & \ArmPower \\

    Mirror substrate & \GwincVal{Blue.Substance} \\

    Mirror radius &
    \SI{\GwincVal{Blue.Materials.MassRadius_cm}}{cm} \\

    Mirror thickness &
    \SI{\GwincVal{Blue.Materials.MassThickness_cm}}{cm} \\

    Beam radius on ITM/ETM\tablefootnote{$1/e^2$ intensity} &
    \num{\GwincVal{Blue.Optics.ITM.BeamRadius_cm}}/\SI{\GwincVal{Blue.Optics.ETM.BeamRadius_cm}}{cm} \\

    ITM transmittance &
    \SI[round-mode=figures,round-precision=3,scientific-notation=true]{\GwincVal{Blue.Optics.ITM.Transmittance}}{} \\

    PRM transmittance &
    \SI[round-mode=figures,round-precision=2,scientific-notation=true]{\GwincVal{Blue.Optics.PRM.Transmittance}}{} \\

    SRM transmittance &
    \SI[round-mode=figures,round-precision=2,scientific-notation=true]{\GwincVal{Blue.Optics.SRM.Transmittance}}{} \\

    Mass per stage &
    50/70/\SusMassPUM{}/\SusMassTM{}\, kg \\

    Final stage temperature & \Tcryo\,K \\

    Final stage construction & \GwincVal{Blue.Substance}
    \GwincVal{Blue.Suspension.FiberType_str} \\

    Final stage length &
    \SI[round-mode=figures,round-precision=2]{\GwincVal{Blue.Suspension.Stage(1).Length}}{m} \\

    Newtonian noise suppression &
    \num{\GwincVal{Blue.Seismic.Omicron}} \\

    Injected squeeze factor &
    \SI{\GwincVal{Blue.Squeezer.AmplitudedB}}{dB} \\

    Squeeze injection loss &
    \num{\GwincVal{Blue.Squeezer.InjectionLoss}} \\

    Squeeze filter cavity length &
    \SI{\GwincVal{Blue.Squeezer.FilterCavity.L}}{m} \\

    Squeeze filter cavity loss\tablefootnote{Round-trip loss; see \cref{sec:filter_cavities}} &
    \SI{\GwincVal{Blue.Squeezer.FilterCavity.Lrt_ppm}}{ppm} \\

    \bottomrule
  \end{tabular}}
  \caption{Relevant parameters for the \voy{} design.}
  \label{tab:params}
\end{table}

%% file: SiliconMasses.tex
\label{s:SiliconMasses}

\subsection{Material}
\begin{figure}
  \centering
  \includegraphics[width=\columnwidth]{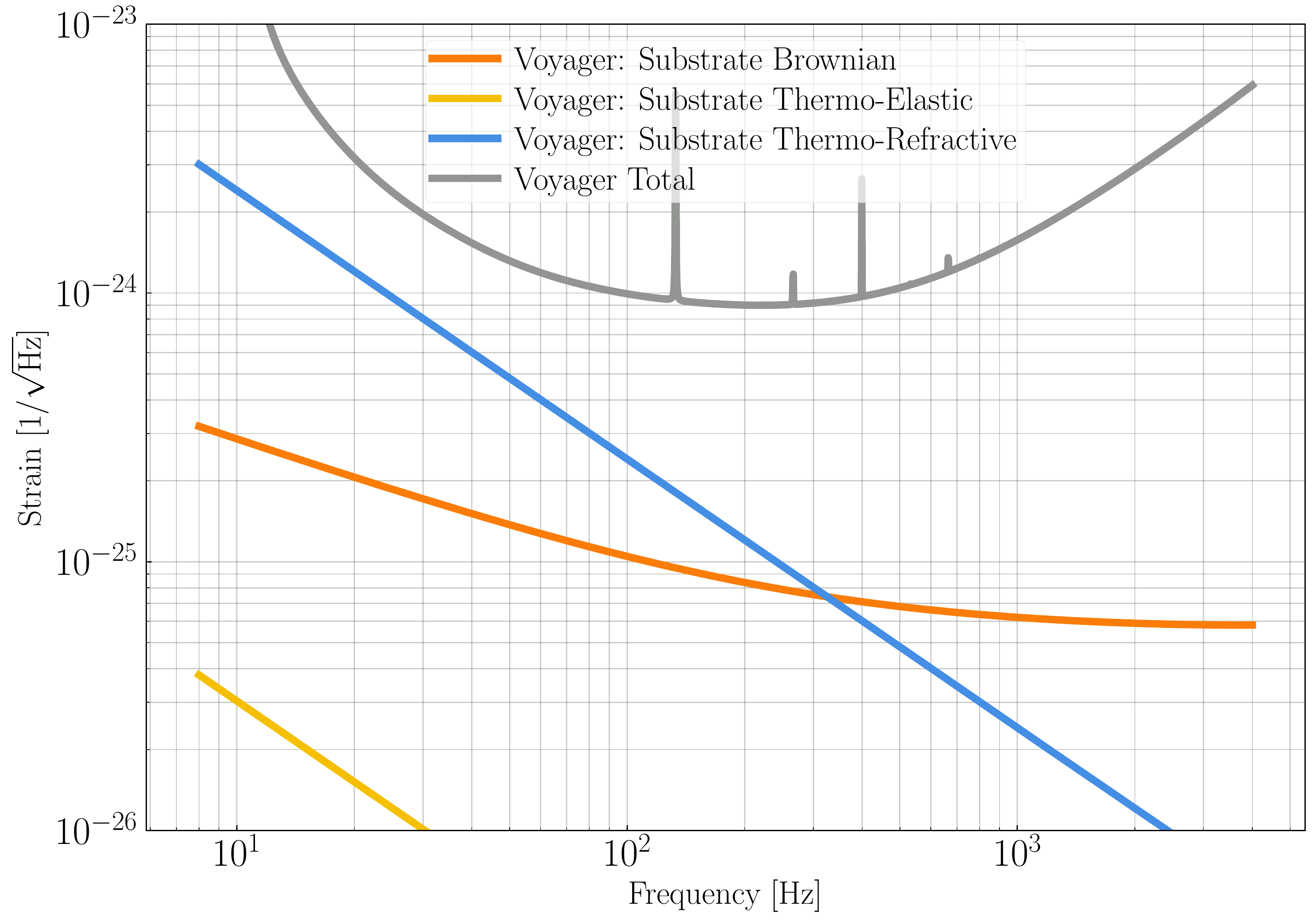}
  \caption{Strain noise from thermally induced noise sources in the \voy \ \Tcryo{}\,K crystalline silicon test masses.  
  }
  \label{fig:SiThermal}
\end{figure}

We have chosen \Tcryo{}\,K crystalline silicon as the test mass material for \voy. \Cref{fig:SiThermal} shows the thermal noise strain curves from crystalline silicon test masses held at \Tcryo{}\,K, where it can be seen that neither Brownian nor thermo-optic substrate noises should limit detector sensitivity. To justify this material and temperature choice, we compare its thermal noise performance with three other materials that are currently used or proposed for use in GW interferometers: fused silica~\cite{aLIGODetectorRef, VirgoDetectorRef}, sapphire~\cite{PhysRevD.89.062003}, and 10\,K silicon~\cite{HiEA2009}.

Thermal noise in a fused silica test mass is limited by Brownian motion, which is related to mechanical loss through the fluctuation-dissipation theorem~\cite{CaWe1951, Kubo:FDT, Callen:1959}.  In fused silica, the mechanical loss has a broad peak below room temperature.  Thus its thermal noise does not benefit from cryogenic cooling~\cite{schroeter2007}. Silicon has lower mechanical loss, and consequently lower Brownian noise than fused silica without any loss peaks at low temperatures~\cite{NumataYamamotoCryogenics}.  %

Sapphire, like silicon, is free of cryogenic loss peaks.  However, thermo-elastic noise is an important noise mechanism in these two crystalline materials, due to their high thermal conductivity.  Thermodynamic fluctuations of heat inside the material are the source of this noise.  The fluctuations are converted to mirror surface displacement through the coefficient of thermal expansion $\alpha$.  The displacement power spectral density is $S_{\rm TE}(f) \sim \kappa \alpha^2 T^2$, where $\kappa$ is the thermal conductivity~\cite{BGV1999,LiTh2000}.

\begin{figure}[ht]
  \centering
  \includegraphics[width=\columnwidth]{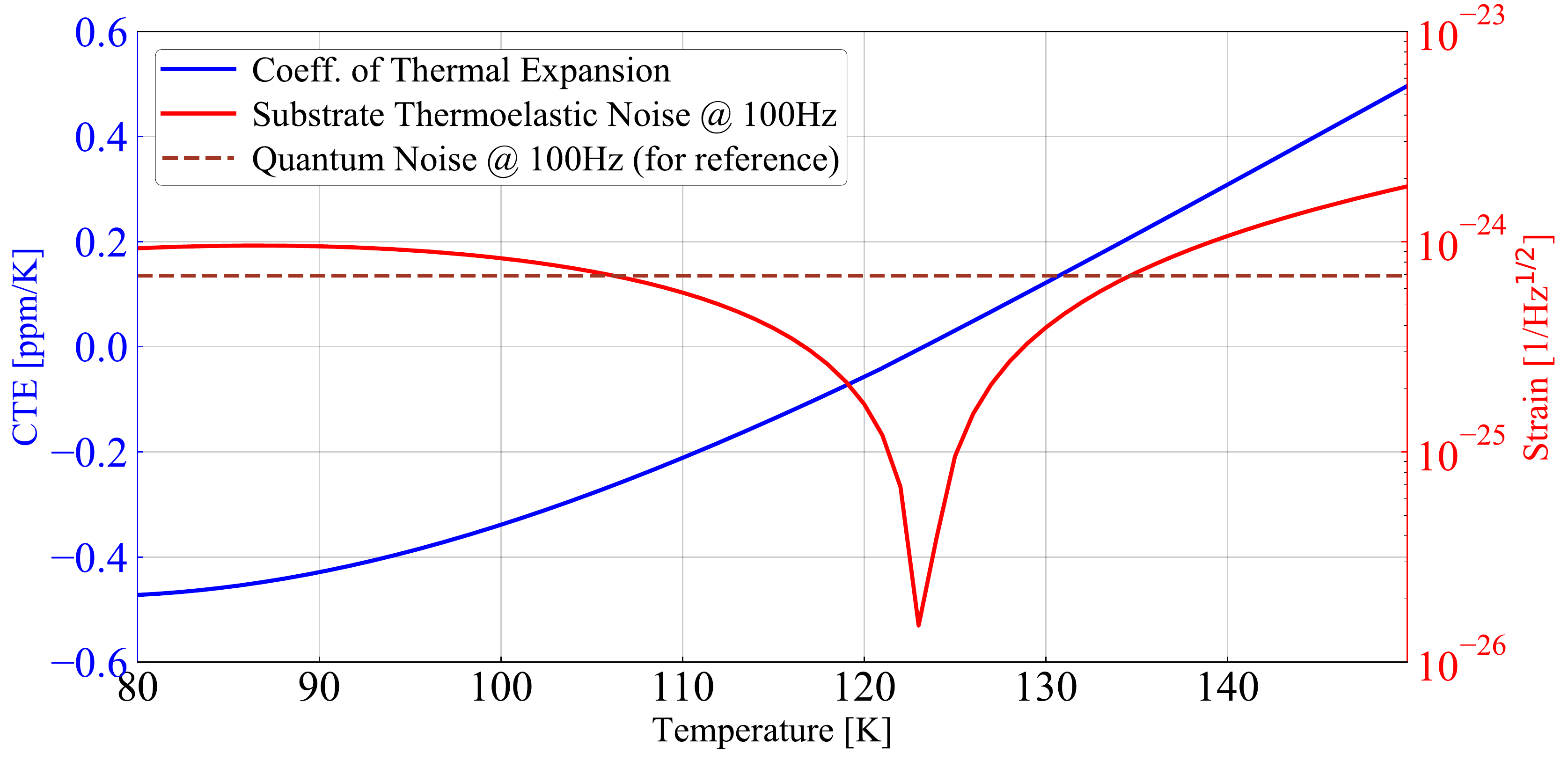}
  \caption{Coefficient of thermal expansion (CTE) vs temperature for silicon (blue). The substrate thermo-elastic noise at 100Hz (solid red) is minimized at 123\,K when the CTE crosses zero~\cite{Kim1992}. Shown for reference is the quantum noise (QN) at 100\,Hz (dashed red), corresponding to a fixed \SI{3}{\mega\watt} in the arms (for simplicity, we have deliberately ignored secondary effects that can cause the stored arm power to vary with temperature, such as temperature-dependent variations in power-handling in the mirrors).}
  \label{fig:magicSi123K}
\end{figure}

Thermo-elastic noise can be mitigated by holding the test mass at a temperature near absolute zero ($\sim$20\,K is sufficient for sapphire), where $\alpha$ must vanish due to the Nernst heat theorem.  Silicon has the unusual property that its thermo-elastic noise is also eliminated at an elevated temperature, \Tcryo{}\,K, where $\alpha$ crosses through zero~\cite{Kim1992} (see \Cref{fig:magicSi123K}).

To operate an interferometer at temperatures in the 10\,--\,20\,K regime requires imposing an austere heat budget on the test mass, which in turn makes it difficult to achieve high circulating power in the arms~\cite{somiya2012detector, HiEA2009}.  By contrast, at \Tcryo{}\,K, the test mass heat budget is compatible with the use of megawatts of circulating power.  
This advance in optical power handling is what will allow us to also reduce the quantum noise, so as to realize the benefit of the improved thermal noise in \Tcryo{}\,K silicon across a broad band of frequencies.

\subsection{Size and composition}

Large, high purity silicon crystals will be required for the \voy{} test masses.

The size of the test mass affects the sensitivity in two ways.  First, larger mirror surfaces enable larger optical spot sizes, thus reducing the coating thermal noise.  Second, heavier masses suffer less disturbance from radiation pressure forces.

Impurities in the silicon can degrade the sensitivity.  The most stringent known requirement derives from the production of free carriers (unbound electrons and holes) by these impurities and, ultimately, the impact this has on the cryogenic cooling system.  To couple light into the arm cavities, a high-power beam must transmit through each input test mass.  Some of the light interacts with free carriers inside the silicon substrate and is absorbed, heating up the test mass.  High purity silicon is needed so that the heating due to free carrier absorption does not exceed the radiative cooling.

The size and composition of silicon crystals available to us are dictated by the commercially viable processes for crystal growth.  Crystals with ultra-low contamination are produced using the float zone technique, but this process has not been scaled up to sizes greater than 20\,cm in diameter.  The magnetically stabilized Czochralski (MCZ) technique, on the other hand, yields 45\,cm crystals that are somewhat less pure than float zone silicon~\cite{SiliconMaterialsSemiconductorHandbook}. MCZ silicon is the most promising candidate for producing test masses of the size needed for \voy.

Oxygen is by far the most abundant impurity in MCZ silicon.  It enters by diffusion from the fused silica crucible that holds the molten silicon, and is typically present at the level of \SI{1E17}{\per\cubic\centi\metre} or even higher.  Most of this oxygen is interstitial to the lattice of silicon atoms, and does not affect the free carrier density.  However, oxygen also forms complexes, referred to as ``thermal donors'', that add free electrons.  Rapid annealing may offer a way to disrupt oxygen complexes and eliminate some of the free carriers they contribute, which would otherwise be the dominant population in undoped MCZ silicon~\cite{OxygenInSiliconChap7, doi:10.1063/1.368586, Kissinger2015}.

Other impurities include carbon, boron, and phosphorus.  Carbon, typically found at \SI{1E15}{\per\cubic\centi\metre}, has little effect on the free carrier density.  Boron and phosphorus are used as dopants to manipulate the carrier density, and they are found even in undoped silicon as contaminants with concentrations $\sim$\SI{1E12}{\per\cubic\centi\metre}.

\subsection{Absorption}
Noteworthy absorption processes in silicon include inter-band absorption, two-photon absorption, and free carrier absorption.
Due to the choice of wavelength and power density in the optics, the inter-band and two-photon absorption are found to be unimportant for \voy~\cite{doi:10.1063/1.4923379, doi:10.1063/1.2737359}.

In the Drude model of free carrier response, the free carrier absorption is calculated as~\cite{Soref1987}:
\begin{equation}
\label{eq:FreeCarrierAbsorption}
\alpha_{\rm FC} = \frac{e^2 \lambda^2}{4 \pi \epsilon_0 n c^3} \frac{n_c}{m_*^2 \mu}
\end{equation}
with $\lambda$ the optical wavelength, $e$ the elementary charge, $n$ the refractive index, $n_c$ the density of free carriers, $m_*$ the carrier effective mass, and $\mu$ the carrier mobility.  (Note that the carrier density, mass, and mobility are different for electrons and holes.)

According to \Cref{eq:FreeCarrierAbsorption}, absorption of roughly 1\,ppm/cm would be expected, if the level of residual boron and phosphorus doping available in MCZ silicon is the limiting factor.
Absorption as low as 4.3\,ppm/cm has been measured at 1550\,nm in float zone silicon~\cite{Degallaix:13}. This result was in excess of the Drude model prediction, possibly due to the existence of an absorption band near 2300\,nm in $n$-type silicon~\cite{PhysRev.108.268}.  Absorption measurements and annealing experiments on MCZ silicon samples are in progress, to better understand the mechanisms that limit absorption, and how thoroughly the contribution of oxygen can be suppressed.

\subsection{Phase noise}
The dominant phase noise term in the substrate is expected to be thermo-refractive noise.  Like thermo-elastic noise, this noise is sourced by thermodynamic fluctuations of heat inside the material.  The fluctuations are converted to refractive index fluctuations through the coefficient $\beta = dn/dT$.  The resulting phase noise is imposed on the light in the signal recycling cavity.  The power spectral density of this noise has been estimated as~\cite{Braginsky:2004fp}:
\begin{equation}
S_{\rm TR}(f) = \frac{4 a \beta^2}{\pi^3 w^4 f^2} \frac{\kappa k_{\rm B} T^2}{\rho^2 C^2}
\end{equation}
in units of signal recycling cavity displacement, for a Gaussian beam of radius $w$, traversing an infinite plate with thickness $a$, where $\rho$ is the density, $C$ is the specific heat capacity, and $\kappa$ is the thermal conductivity.  For \voy, thermo-refractive noise is expected to be below the coating and quantum noise terms, but still within a factor of a few of limiting the sensitivity, as shown in \Cref{fig:SiThermal}.

Analogously, the density of free carriers in silicon has an effect on the refractive index, so that thermodynamic fluctuations of the carrier density $n_{\rm c}$ impose phase noise.  The magnitude of this effect is described by a carrier dispersion coefficient $\gamma_{\rm c} = dn/dn_{\rm c}$ (different for electrons and holes).  The carrier density noise was estimated as~\cite{CDNoiseBruns2020}:
\begin{equation}
S_{\rm CD}(f) = \frac{2 n_{\rm c} \gamma_{\rm c}^2 a l_{\rm D}^2}{\pi w^2 D_{\rm c}}
\end{equation}
referred to signal recycling cavity displacement, where $D_{\rm c}$ is the carrier diffusion coefficient (also different for electrons and holes), and $l_{\rm D}$ is the Debye length. Although this noise has yet to be experimentally validated, the noise level was estimated to be less than $10^{-28} /\sqrt{\rm Hz}$, and thus is expected to be negligible for \voy.

\subsection{Scattering}
The absorption, refractive index, birefringence, and surface profile of the test masses should all be uniform spatially, as far as possible.  Any spatial inhomogeneity leads to scattering of the light that interacts with the test mass.  Scattering is problematic because the loss of light can limit the buildup of optical power in the cavities.  Even worse, scattered light often finds a path to return to the interferometer, thus contaminating the output with the ambient noise of all surfaces it encountered along the way.

The specific requirements on these characteristics will be determined as part of the detailed optical design of \voy{}.  The tolerable amount of scattered light will be smaller than specified for Advanced LIGO~\cite{T000127}.  However, \voy{} will also be less prone to wide-angle scattering, as discussed in \cref{s:scatter}.

Spatial gradients in the atomic impurities discussed above are one likely source of inhomogeneity in MCZ silicon crystals.  Another is microscopic crystal defects, such as voids, stacking faults, and SiO$_2$ precipitates~\cite{Vanhellemont2015}.

Impurity and defect populations can be manipulated during the crystal growth process, and also to some extent by annealing of the finished crystal.  If we suppose that voids are the predominant defect population, approximated as spheres with a characteristic radius of 100\,nm, then we can compute their scattering cross-section due to Mie scattering at wavelength 2000\,nm.  For a void concentration of \SI{1e3}{\per\cubic\centi\metre}, the resulting loss is estimated as 10\,ppm per round trip through a \voy{} test mass.  Measurements to check the level of scatter loss in MCZ silicon crystals are underway~\cite{G1700998}.

\subsection{Thermal lensing and active wavefront control}
\label{s:thermalLensing}
GW interferometers suffer from the detrimental effects of thermal gradients and distortion due to absorption of optical power~\cite{Brooks:16,Winkler:TCS} in the surface and substrates of the core optics. \voy{} is no exception, but the high thermal conductivity of silicon at cryogenic temperatures helps to mitigate this issue.

Analogous to the Advanced LIGO thermal compensation system \cite{Brooks:16}, in \voy{} there are two room-temperature silicon compensation plates in the recycling cavities, as illustrated in \Cref{fig:IFO_schematic}, to which thermal actuation can be applied to correct for lensing in the substrates of the core optics. Room-temperature silicon is preferable to fused silica for the compensation plates for two reasons:
\begin{itemize}
    \item Silicon's thermal lensing per watt is a factor of six greater at \SI{300}{\kelvin} than at \SI{123}{\kelvin}.  Consequently there is much more actuation per watt in the compensation plate than distortion per watt in the test mass, yielding a comfortable measure of control on the thermal lensing.
    \item Due to the increased absorption of mid-IR wavelengths (particularly at longer wavelengths), self-heating in a fused silica compensation plate may produce a larger thermal lens than the one to be corrected in the test mass (see \Cref{s:fused_silica_absorption} for details). Absorption in the room-temperature silicon compensation plates is expected to be comparable to that found in the test masses~\cite{PhysRev.108.268, Degallaix_2014}.
\end{itemize}

Point absorbers on the reflective surface of the test mass have impaired the performance of Advanced LIGO~\cite{PointAbsorbers2019}. However, \voy{} will not suffer from this problem, as the coefficient of thermal expansion of the silicon test mass is effectively zero at \SI{123}{\kelvin}. The surface deformation from point absorbers at full operating power is expected to be at least $1000$ times smaller than in Advanced LIGO.

Finally, the current design has not specified a way to tune the radius of curvature of the test masses (in Advanced LIGO this tuning relies on a non-zero coefficient of thermal expansion).  
Unless such an actuator can be devised, the curvature error tolerance will be  tighter than in Advanced LIGO. 
The curvature tolerance  will be computed using a full simulation/model of the interferometer that includes the effects of control loops and higher order modes. This is still under development and beyond the scope of this design paper. However, we indicate here the considerations that will impact the tolerance specification: 
\begin{itemize}
\item optimizing the mode-matching between the two arm cavities (to minimize the differential loss),
\item optimizing the mode-matching to the two recycling cavities (to maximize power build up, signal bandwidth and squeezing efficiency),
\item ensuring the overall design of the arms is such that no higher order modes are close to resonant in the arms, and
\item ensuring the arm cavities are designed to minimize the number of parametric instabilities that have to be damped, see \Cref{s:PI}.
\end{itemize}

%% file: Coatings.tex
\label{s:Coatings}

Gravitational wave interferometers use
ion beam sputtered (IBS) thin films as the high reflectivity (HR)
optical coatings on the test masses. These films are made of
amorphous oxides~\cite{Reid:2016wg}. After decades of development,
these optical coatings now have excellent optical properties.
Unfortunately, their internal friction (mechanical loss) is still large, and therefore, the concomitant Brownian noise
is the limiting displacement noise in the Advanced LIGO design in the
40\,--\,200\,Hz frequency band  \cite{Massimo2016, Pinard:17}.

Absorption of the interferometer beam in IBS coatings is problematic for interferometer operation because the resulting thermo-optic lenses alter the optical configuration from the nominal design. This problem has been present in all previous interferometer designs, and \voy{} is no exception (see \Cref{s:thermalLensing} for more details). However, optical absorption poses a special challenge for \voy{}: when the heat load from absorbed optical power is coupled with cryogenic interferometer operation, a limit is placed on the maximum power stored in the interferometer.

This section describes the design of a cryogenic, amorphous-silicon-based, IBS coating for \voy{} that decreases  the coating Brownian noise by a factor of 4 compared to Advanced LIGO. In \Cref{s:Brownian}, we discuss the remarkably low mechanical loss of amorphous silicon at \SI{123}{K} that makes this noise reduction possible.  As it is relevant to the overall heat budget, the current state of absorption in amorphous silicon coatings is considered in \Cref{s:opticalAbsorption}.

\subsection{Basic optical requirements}
\label{s:CoatReqs}

We begin with a brief review of the basic requirements for the test mass HR coatings. 

\begin{itemize}
\item \TMDiameterCm{} diameter: the coatings must extend across the full diameter of the silicon test masses.
\item High reflectivity on ETM (Transmittance T = \SI[round-mode=figures,round-precision=3,scientific-notation=true]{\GwincVal{Blue.Optics.ETM.Transmittance}}{})
\item High reflectivity on ITM (T = \pgfmathparse{round(10000*\GwincVal{Blue.Optics.ITM.Transmittance})/100}\pgfmathresult\%)
\item Low scatter loss ($\le$ \SI[round-mode=figures,round-precision=3,scientific-notation=true]{\GwincVal{Blue.Optics.Loss}}{} per bounce)
\item Cancellation of thermo-optic noise \cite{Evans2008,Hong2013}.
\item Reduction of Brownian noise by a factor of 4 or 5 from Advanced LIGO levels
\item At most 1ppm absorption, set by the heat budget of the test mass, see \Cref{s:Cryo}.
\end{itemize}

\subsection{Brownian noise}
\label{s:Brownian}

\begin{table}[tbhp]
\centering
\begin{tabular}{lcccc} \toprule 
Parameter & Detector & Material & Loss-angle & Refractive Index \\ 
 & & & ($\phi$) & $n$ \\ \midrule
Low index & aLIGO (300\,K) & SiO$_2$ & $4.0\times10^{-5}$ & 1.45 \\ 
High index & aLIGO (300\,K) & Ta$_2$O$_5$ & $2.3\times10^{-4}$ & 2.07 \\ \midrule
Low index & Voyager (123\,K)& SiO$_2$ & $1.0\times10^{-4}$ & 1.436 \\ 
High index & Voyager (123\,K)& $\alpha$-Si & $\le1.0\times10^{-5}$\cite{Hellman:2014} & 3.5 \\ \bottomrule
\end{tabular}
\caption[Summary of the coating material parameters]
        {Summary of the coating material parameters. Note that, due to the peculiarities of glass,  the loss-angle for the SiO$_2$ increases at cryogenic temperatures\cite{Martin2014}.}
        \label{tab:coating_parameters}
\end{table}

As described in \cite{Evans2008}, Brownian noise in the coating is the dominant residual noise source, particularly when thermo-optic noise is minimized. Brownian noise is driven by  mechanical dissipation, where the relation between the dissipation and the noise is described by Callen's Fluctuation-Dissipation Theorem~\cite{CaWe1951, Kubo:FDT, Callen:1959}:
\begin{equation}
S_x(f) = \frac{k_B T}{\pi^2 f^2} \left| Re \big[ Y(f) \big]\right|,
\label{eq:FDT}
\end{equation}
where $S_x(f)$ is defined as the power spectral density of physical quantity $x$.
$T$ is the temperature of the mirror and $Y(f) \equiv \dot{x}(f)/F(f)$ is the complex mechanical admittance (the inverse of the mechanical impedance) associated with the radiation pressure force of the Gaussian intensity profile laser beam. 

For a single layer coating, it can be shown that the Brownian noise spectrum is proportional to the mechanical loss angle, $\phi$, of the layer. The Brownian noise of a multi-layer coating will involve the complex weighted sum of all the loss angles of all the layers.

\subsubsection{Amorphous Silicon}
Although almost all amorphous thin films suffer from a high level of internal friction, there is a film that has been made with nearly no such loss: amorphous silicon ($\alpha$-Si)~\cite{Hellman:2014, Pohl:RMP}. Recent measurements~\cite{Reid2018a} have shown that amorphous silicon can be grown with both very low mechanical loss and low optical absorption at \SI[round-mode=figures]{2}{\micro\meter}. \Cref{tab:coating_parameters} compares the loss angles for the Advanced LIGO and Voyager coating materials. Note that the loss angle, $\phi$, for $\alpha$-Si is more than a factor of 20 lower than the high index material used in Advanced LIGO.

\begin{figure}
\centering
\begin{subfigure}[b]{0.75\textwidth}
   \includegraphics[width=1\linewidth]{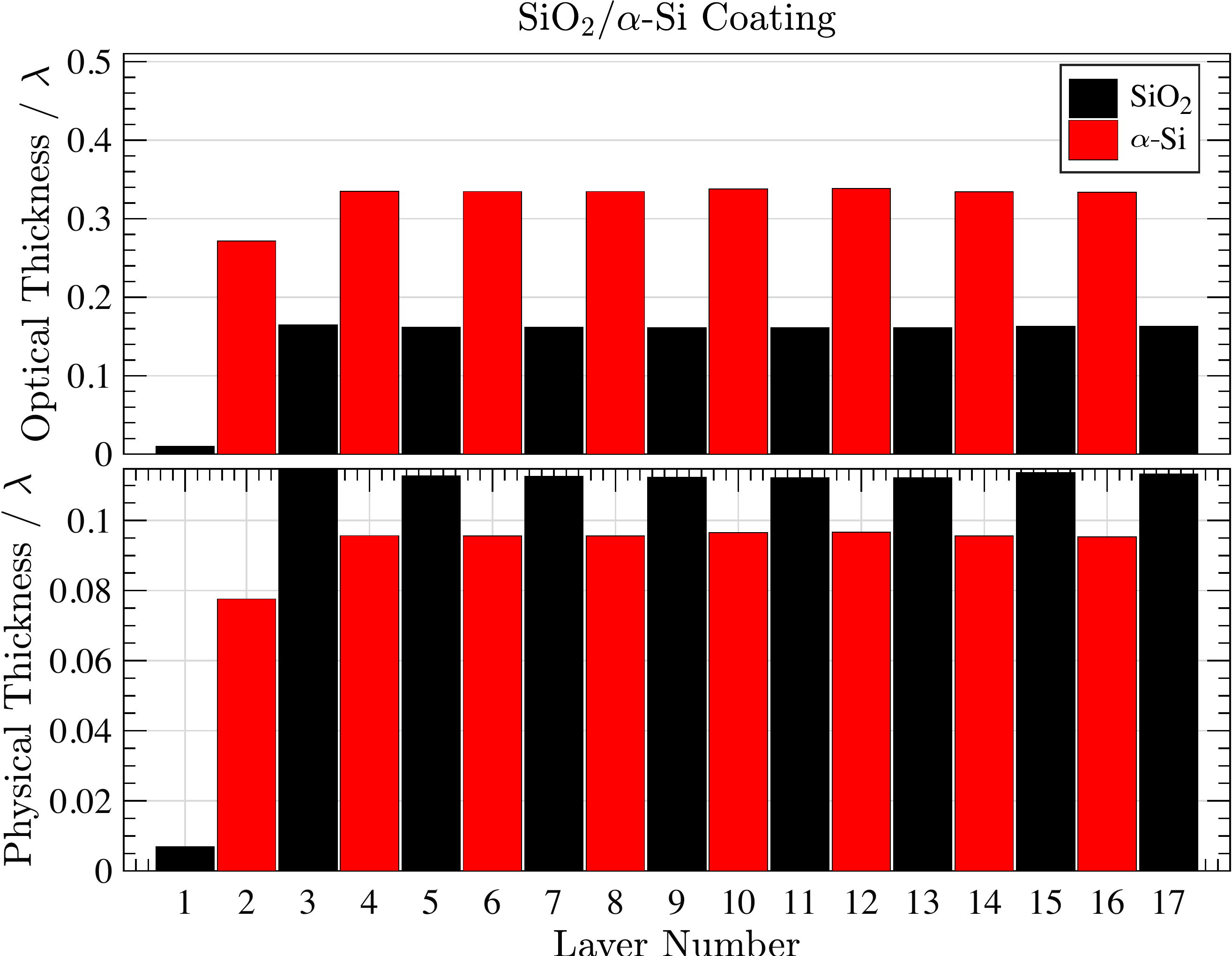}
   \caption{}
   \label{fig:aSiCoatingDesign} 
\end{subfigure}

\begin{subfigure}[b]{0.75\textwidth}
   \includegraphics[width=1\linewidth]{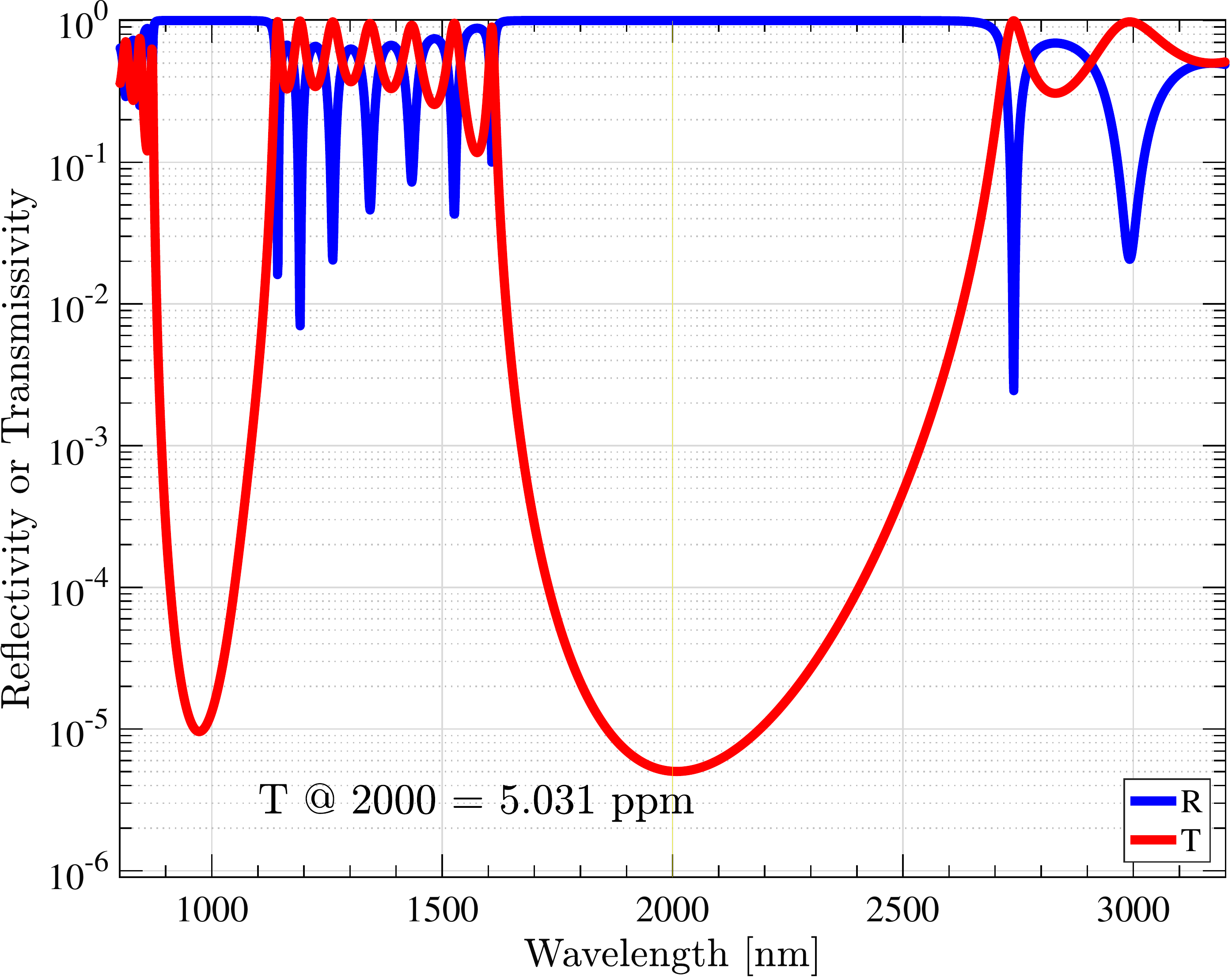}
   \caption{}
   \label{fig:aSiCoatingSpectrum}
\end{subfigure}
\caption{(a) Layer structure for the 
    $\alpha$-Si:SiO$_2$ HR coating for end test masses. This coating design was optimized to minimize Brownian noise, meet the 5\,ppm transmission goal, and minimize first order sensitivity to coating thickness and index of refraction errors. (b) Reflection and transmission calculations
    for the $\alpha$-Si:SiO$_2$ HR coating.}
\end{figure}

\begin{figure}[h]
  \centering
  \includegraphics[width=\columnwidth]{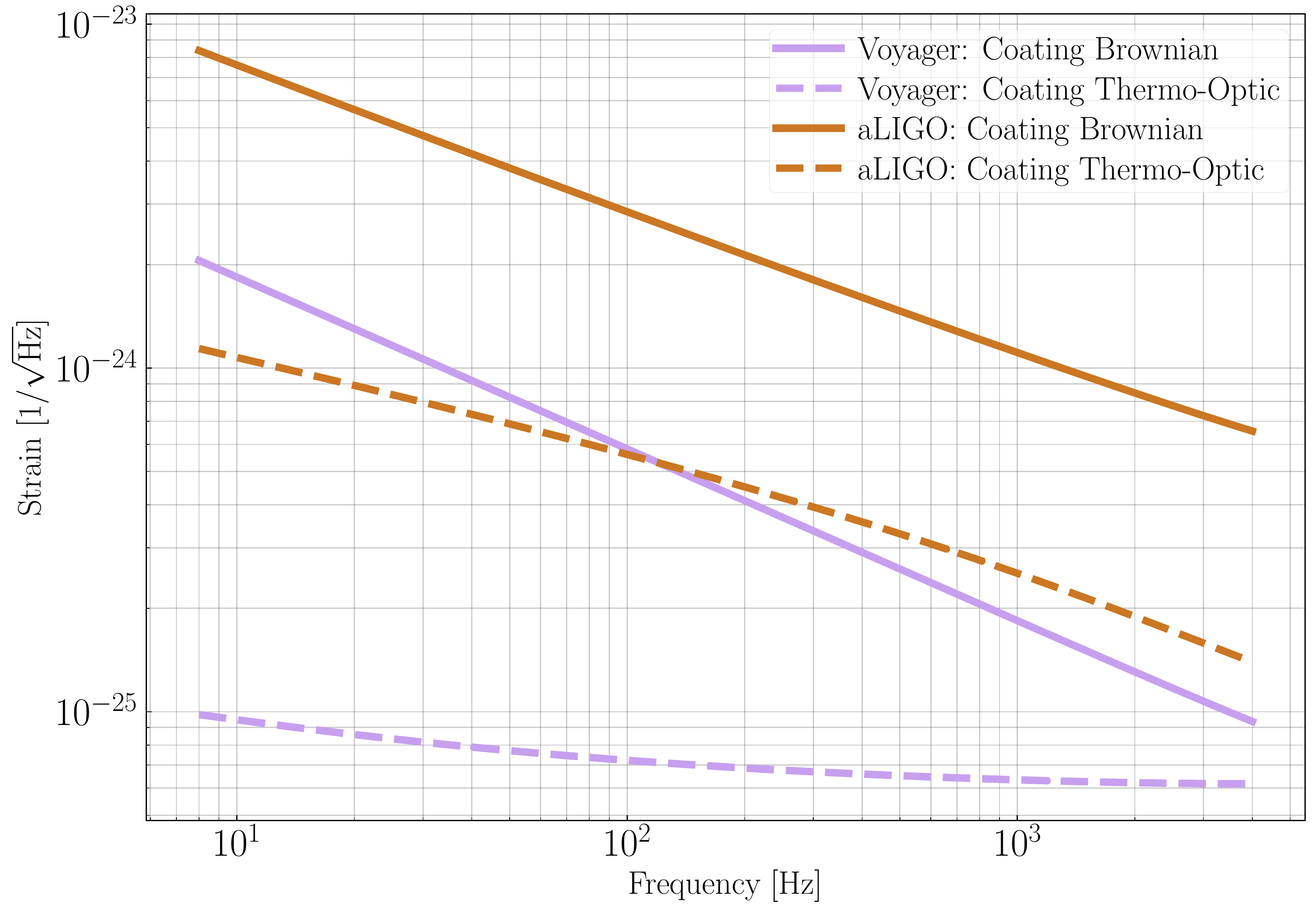}
  \caption[$\alpha$-Si coating noise]{Mirror thermal noise sources for the LIGO Voyager and Advanced LIGO.}
  \label{fig:aSiCoatingNoise}
\end{figure}

Using the material parameters for $\alpha$-Si:SiO$_2$ at \Tcryo\,K{} found in the literature, we have numerically optimized the layer structure so as to minimize the overall displacement noise while maintaining a low sensitivity to layer thickness variations (details of this technique can be found in \cite{Hong2013}). The result is an ETM coating with 5\,ppm transmission. \Cref{fig:aSiCoatingDesign} shows the coating structure (notice that the design is close to, but not exactly, a simple stack of layers of $\lambda/4$ thickness). The transmission and reflection spectra are shown in \Cref{fig:aSiCoatingSpectrum}. Finally, \Cref{fig:aSiCoatingNoise} shows the Brownian and thermo-optic noises for Advanced LIGO and \voy{}; Brownian noise is the limiting coating noise source for both, but it is more than 4 times lower for Voyager compared to aLIGO.

It is noteworthy that, unlike in today's gravitational wave detectors, the contribution to the Brownian noise from the high refractive index ($\alpha$-Si) layers is so small that the low index (SiO$_2$) layers become the dominant contributor to the noise.

\subsubsection{Crystalline coatings}
\label{s:crystalCoatings}

Crystalline coatings such as AlAs:GaAs~\cite{Cole:2016ud} and
AlGaP:GaP~\cite{Angie:2013} have been shown to have a higher mechanical
Q than amorphous dielectric coatings and, as such, are a favorable technology to pursue for high
precision optical cavities. The thermo-optic noise of these coatings is generally high, but it can be mitigated by careful design of layer thicknesses~\cite{Chalermsongsak_2016}. Both crystalline coating options show promise as candidates for \voy{} but require significant further development. AlGaP:GaP is lattice matched to silicon and could therefore be epitaxially grown directly onto a test mass substrate, but the absorption must be reduced to the 1\,ppm level. AlAs:GaAs is not lattice matched to silicon so must be grown on a GaAs substrate and then lifted off~\cite{Yablonovitch1990} and affixed to the silicon test mass face, a technique yet to be demonstrated for 30\,cm diameter coating stacks. While an $\alpha$-Si:SiO$_2$ coating is the current choice for \voy{}, breakthrough results on crystalline coatings could lead to a switch in design.

\subsection{Optical absorption}
\label{s:opticalAbsorption}

The design of \voy{} allows for 1\,ppm absorption in the coatings of the test masses, stemming from the need to keep the core optics at cryogenic temperatures (see \Cref{s:Cryo}). Much research has been performed in the last few years with the aim of lowering the optical absorption, although an $\alpha$-Si coating with absorption of less than 1\,ppm at \SI{2000}{\nano\meter} and \Tcryo\,K has yet to be demonstrated.  However, it appears likely this can be achieved, based on two recent results:
\begin{itemize}
    \item The absorption in $\alpha$-Si coatings was consistently measured to be approximately 7 times lower at \SI{2000}{\nano\meter} than  \SI{1550}{\nano\meter} \cite{Steinlechner:2017}, and also improves with cooling.
    \item Using a novel ion-beam deposition method, Birney et al.~\cite{Birney:2017dt} were able to produce an $\alpha$-Si coating with absorption of 7.6\,ppm at \SI{1550}{\nano\meter} and room temperature.
\end{itemize}

\noindent Taken together, these two results suggest that an $\alpha$-Si coating with less than 1\,ppm absorption is feasible.

%% file: Wavelength.tex
\label{s:Wavelength}
The \voy{} design uses silicon test masses, which are effectively opaque for wavelengths shorter than approximately \SI{1100}{\nano\meter}.
Thus, the laser wavelength used in first and second generation GW detectors (\SI{1064}{\nano\meter}) will not work in \voy{}, and a new laser wavelength must be chosen.  

As discussed in \Cref{s:Lasers}, we require approximately \LaserPowerApprox{} of single-frequency laser power at the input to the interferometer. These requirements (high power and low noise) demand a mature CW laser source.
Within the 1400\,--\,\SI{2100}{\nano\meter} range, the main laser sources are telecommunication lasers at \SI{1550}{\nano\meter} and thulium- and holmium-based sources in the 1800\,--\,\SI{2100}{\nano\meter} band.
These are the wavelengths we will consider.

The laser wavelength affects the performance of virtually all the optical elements in the interferometer, many of which will directly impact the interferometer sensitivity, as discussed in the remainder of this section.  Although many considerations enter into the choice between 1550 and \SI{2000}{\nano\meter}, the decisive factor is the absorption in the mirror coatings.  Selecting a longer wavelength, around \SI{2000}{\nano\meter}, appears to be necessary in order to achieve the designed arm cavity power in \voy{}, with other side effects being of secondary importance.

To justify a choice of wavelength, this section collects and discusses different physical processes (photodiode QE, coating absorption, substrate absorption, etc)  that are discussed elsewhere in this manuscript. This section explores impacts solely with respect to wavelength; other sections explore these concepts individually and in a more multi-faceted way.

\subsection{Quantum limits}
For a fixed arm cavity power, the shot noise limited strain sensitivity at high frequencies degrades proportionally with the square root of the laser wavelength.
Conversely, the radiation pressure limited strain sensitivity at low frequencies improves with increasing wavelength. 
From a quantum noise standpoint, increasing the laser wavelength by a factor of 2 is equivalent to lowering the arm cavity power by a factor of 2,  all else being equal.  However, the available arm cavity power is also constrained by other factors, primarily the coating absorption effect discussed above.

\subsubsection{Photodetector quantum efficiency}
\label{s:QEpre}

High photodetector quantum efficiency (QE) is essential to make good use of high levels of squeezing.
$\mathrm{QE} > 99\%$ will be required for \voy{}.  At the time of writing, the QE of InGaAs photodetectors at \SI{1550}{\nano\meter} is already sufficient to meet this requirement~\cite{Mehmet:2011je}. 
At \SI{2000}{\nano\meter}, $\mathrm{QE} \gtrsim 90\%$ has yet to be demonstrated for InGaAs.
Currently, \SI{1550}{\nano\meter} is a better choice of wavelength from the perspective of QE.
However, we know of no fundamental obstacle to achieving near-unity QE in photodetectors around \SI{2000}{\nano\meter}.
Photodetectors for \SI{2000}{\nano\meter} are discussed in more detail in \Cref{s:QE}.

\subsection{Noise Sources}
\subsubsection{Coating thermal noise}
The coating layer structure and thickness depend upon the wavelength.  In general, a longer operating wavelength requires a proportionally thicker coating, and so the coating thermal noise increases roughly as the square root of the wavelength.  This implies a $\sim 14\%$ degradation in coating thermal noise at 2000\,nm, relative to 1550\,nm.

Amorphous silicon remains the best coating material available for NIR operations (from a thermal noise standpoint); however, the low index bilayer's performance could be improved by changing from SiO$_2$ to either alumina (Al$_2$O$_3$) or SiN which do not have the low temperature mechanical loss peaks.

\subsubsection{Optical scatter loss and noise}
\label{s:scatter}
For a mirror with a given roughness, the total power scattered into wide angles scales as $1/\lambda^2$~\cite{adhikari2019integrated}. We expect approximately 66\% more loss via wide-angle scattering from \SI{1550}{\nano\meter} vs.~\SI{2000}{\nano\meter}
\footnote{It is assumed that the roughness of the mirror coating is independent of the detailed coating layer structure}.
Advantages of reducing the scatter loss include:
\begin{itemize}
\item a higher power recycling gain due to lower loss in the arm cavities
\item lower loss in the high-finesse, squeezing filter cavity
\item reduced backscattering noise (currently limiting all ground based detectors)
\end{itemize}

These in turn lead to reduced requirements on the input laser power, the length of the filter cavity, and scattered light beam baffles, respectively.

\subsubsection{Residual gas noise}  
The phase noise due to residual gas in the main beam tubes~\cite{Zucker:Gas, TAMA:gas} is mainly due to H$_2$ and N$_2$, which have a negligible wavelength dependence in the NIR band.
At atmospheric pressure, there are wide absorption bands~\footnote{\url{https://www.gemini.edu/sciops/telescopes-and-sites/observing-condition-constraints/ir-transmission-spectra}}
near 2000\,nm due to water vapor. At UHV pressures, however, it can be assumed that there is no broadening of resonance linewidths due to particle collisions, but the distribution of particle velocities will create a Doppler resonance profile. The measured pressure for H$_2$O in the LIGO beamtubes is 10$^{-10}$\,Torr; at this level any particular resonances can be avoided by tuning the main laser frequency by several GHz.

The atmospheric absorption is not an issue for the main interferometer, but could be an issue for some of the high power, in-air, laser systems. This issue would drive the laser wavelength higher (e.g. to 2128\,nm) to where the absorption is minimal.

\subsection{Absorption and Impact on Cryogenics}
\subsubsection{Absorption in the HR coatings}
At \Tcryo{}\,K, radiative cooling can extract at most \SI{10}{\watt} of heat from the test masses, as described in \Cref{s:Cryo}. To keep the heat budget in balance, we can tolerate no more than \SI{3}{\watt} of absorbed power in the coating. With \SI{3}{\mega\watt} incident on the optical surfaces,
absorption in the coatings must be very low ($\lesssim 1$ppm) in order to maintain cryogenic temperature. 

Measurements of absorption in amorphous silicon coatings show strong wavelength dependence, with the absorption being much higher at \SI{1550}{\nano\meter} than at \SI{2000}{\nano\meter} \cite{Steinlechner:2017}. The physical mechanism for this is not well understood. However, at present it appears that \SI{2000}{\nano\meter} will be the superior choice of wavelength to reach the objective of high power cryogenic operation.

\subsubsection{Absorption in the test mass substrate}
Substrate absorption is largely determined by the purity of the silicon material and its thermal history, as described in \Cref{s:SiliconMasses}.
According to \Cref{eq:FreeCarrierAbsorption}, the absorption is expected to scale with $\lambda^2$, being $\sim66\%$ higher at \SI{2000}{\nano\meter} than at \SI{1550}{\nano\meter}.
Substrate absorption is an important component of the heat budget for the input test masses, and the arm cavity finesse in \voy{} will be substantially higher than in Advanced LIGO in order to manage this heat source. With the nominal design parameters (cf.~\Cref{tab:params}), the heat load in the substrate of the ITM will be about a factor of three less than that due to the coating absorption. This ultimately drives the design towards longer wavelengths.

\subsubsection{Absorption in auxiliary fused silica components}
\label{s:fused_silica_absorption}
Fused silica will likely be the substrate material for all optics other the test masses and compensation plates.
Absorption of optical power in fused silica, in the absence of OH in the glass, still occurs due to an intrinsic multi-phonon absorption process associated with the Si-O bonds in fused silica. This shows a large increase in absorption around \SI{2000}{nm}~\cite{Tropf95, Kitamura:07}. Absorption of optical power in these optics, most notably the beam-splitter (BS), will cause thermal lensing and loss of power without mitigation by thermal compensation~\cite{Brooks:16}.
For comparison, the estimates of the theoretical limits for absorption in fused silica~\cite{thomas_optical_2006} are approximately:
\begin{itemize}
    \item $<$\,1\,ppm/cm at \SI{1550}{\nano\meter}, ($\approx$ \SI{0.03}{\watt} absorbed in BS)
    \item 20\,ppm/cm at \SI{1900}{\nano\meter}, ($\approx$ \SI{0.6}{\watt} absorbed in BS)
    \item 40\,ppm/cm at \SI{2000}{\nano\meter}, ($\approx$ \SI{1.1}{\watt} absorbed in BS)
    \item 90\,ppm/cm at \SI{2100}{\nano\meter}, ($\approx$ \SI{2.5}{\watt} absorbed in BS)
    \item 120\,ppm/cm at \SI{2128}{\nano\meter}, ($\approx$ \SI{3.3}{\watt} absorbed in BS)
\end{itemize}
versus $<$\,0.06\,ppm/cm at \SI{1064}{\nano\meter} 
(where the BS is \SI{9}{\centi\meter} thick and the substrate sees half of the \BSPower{} in the PRC). The elevated absorption at the longer end of the wavelength range could present significant engineering challenges (strong thermal lenses, increased losses, power imbalance between the arms leading to increased technical noise couplings, increased contrast defect, etc).  It may be possible to decrease the absorption by transitioning to glass made of a material with a heavier molecular mass, such as fluoride \cite{Lines1998}. The technical challenges presented by   wavelength dependent absorption are an active area of research requiring a full interferometer model to analyze the effects in a quantitative way. The results of this will impact the final choice of wavelength. 

The absorption in fused silica opens up an intriguing prospect for an alternative thermal compensation design. Recent work in optical fibers \cite{Dragic:17} has demonstrated that by doping SiO$_2$ with P$_2$O$_5$, which has a negative thermo-refractive coefficient equal to
\SI[round-mode=figures,round-precision=3,scientific-notation=true]{-13.3E-6}{\per\kelvin}, it is possible to tune the $dn/dT$ of the resulting phosphosilicate glass. 
If we were to use fused silica compensation plates (instead of room-temperature silicon), the absorption of the interferometer laser in the glass coupled with a precisely tuned $dn/dT$ could be made to significantly cancel the thermal lens in the substrate of the test mass, thereby rendering the interferometer (mostly) thermally self-correcting.

\subsection{Radiation Pressure Instabilities}

\subsubsection{Opto-Mechanical Angular Instability}
\input{SiggSidles}

\subsubsection{Parametric Instabilities}
\input{PI}

\subsection{Summary}

The wavelength considerations for \voy{} are summarized in \Cref{tab:wavelength_summary}. The color scheme varies through red, orange,  yellow and green corresponding to a variation from negative  to positive situations. As stated at the top of this section, and visually indicated in this table, the coating absorption favors a longer wavelength (around \SI{2000}{\nano\meter}), with absorption in fused silica potentially excluding longer wavelengths if it cannot be mitigated.

\begin{table}[b]
\centering
\begin{tabular}{lcccc} \toprule
{Consideration} &\multicolumn{4}{c}{{Wavelength}}  \\
& \SI{1550}{\nano\meter} & \SI{1900}{\nano\meter} & \SI{2000}{\nano\meter} &   \SI{2128}{\nano\meter} \\ \midrule
Photodiode Q.E. &  \cellcolor{GoodGreen} $>99\%$ & \multicolumn{3}{c}{ \cellcolor{NotSureOrng}  $\approx87$\%. Promising trajectory (\Cref{s:QE}). }  \\
Coating thermal noise & \cellcolor{GoodGreen} Low & \multicolumn{3}{c}{ \cellcolor{OKYellow} $\approx$14\% larger }  \\
Optical scatter loss & \cellcolor{NotSureOrng} 66\% larger & \multicolumn{3}{c}{ \cellcolor{GoodGreen}   Low }  \\
Residual gas noise & \cellcolor{GoodGreen} low H$_2$O  & \cellcolor{NotSureOrng} some H$_2$O & \multicolumn{2}{c}{ \cellcolor{GoodGreen} low H$_2$O }   \\
Coating absorption & \cellcolor{BadRed} High  & \multicolumn{3}{c}{ \cellcolor{OKYellow} Medium } \\
Si substrate absorption &\multicolumn{4}{c}{\cellcolor[rgb]{0.75 0.75 0.75} Increases as $\lambda^2$ but not dominant effect} \\
SiO$_{2}$ substrate absorption  & \cellcolor{GoodGreen} $<1$ ppm/cm & \cellcolor{OKYellow} 20 ppm/cm & \cellcolor{NotSureOrng} $40$ ppm/cm & \cellcolor{BadRed}120 ppm/cm  \\
Angular instability &  \cellcolor{OKYellow} Less stable & \multicolumn{3}{c}{ \cellcolor{GoodGreen}   More stable arm cavity }   \\
Parametric instability &\multicolumn{4}{c}{\cellcolor[rgb]{0.75 0.75 0.75} Very little change with wavelength}  \\ \bottomrule
\end{tabular}
\caption[Summary of wavelength considerations]
{Summary of wavelength considerations}  \label{tab:wavelength_summary}
\end{table}

%% file: SiggSidles.tex
\label{s:SiggSidles}
Optical power, circulating in the arm cavities, applies torque on the 
mirrors and changes the dynamics of the 
suspended mirrors~\cite{Sidles:2006un, Hirose:10, Dooley:ASC:2013}. 

The magnitude of this radiation pressure induced optical torque depends 
upon the optical power and $g$-factors of the cavities. The circulating power 
acts as a spring with either positive or negative stiffness. The sign of 
the feedback depends on the misalignment mode. In the case when two test 
masses have equal radius of curvature, a tilt of the axis produces a restoring 
torque; if the optical axis shifts, then radiation pressure torque tends to further misalign the mirrors. In one case the torque induced by radiation 
pressure makes the suspension mode stiffer (hard), while in the other 
case it tends to make the mode less stiff (soft).

\begin{figure}
  \centering
  \includegraphics[width=\columnwidth]{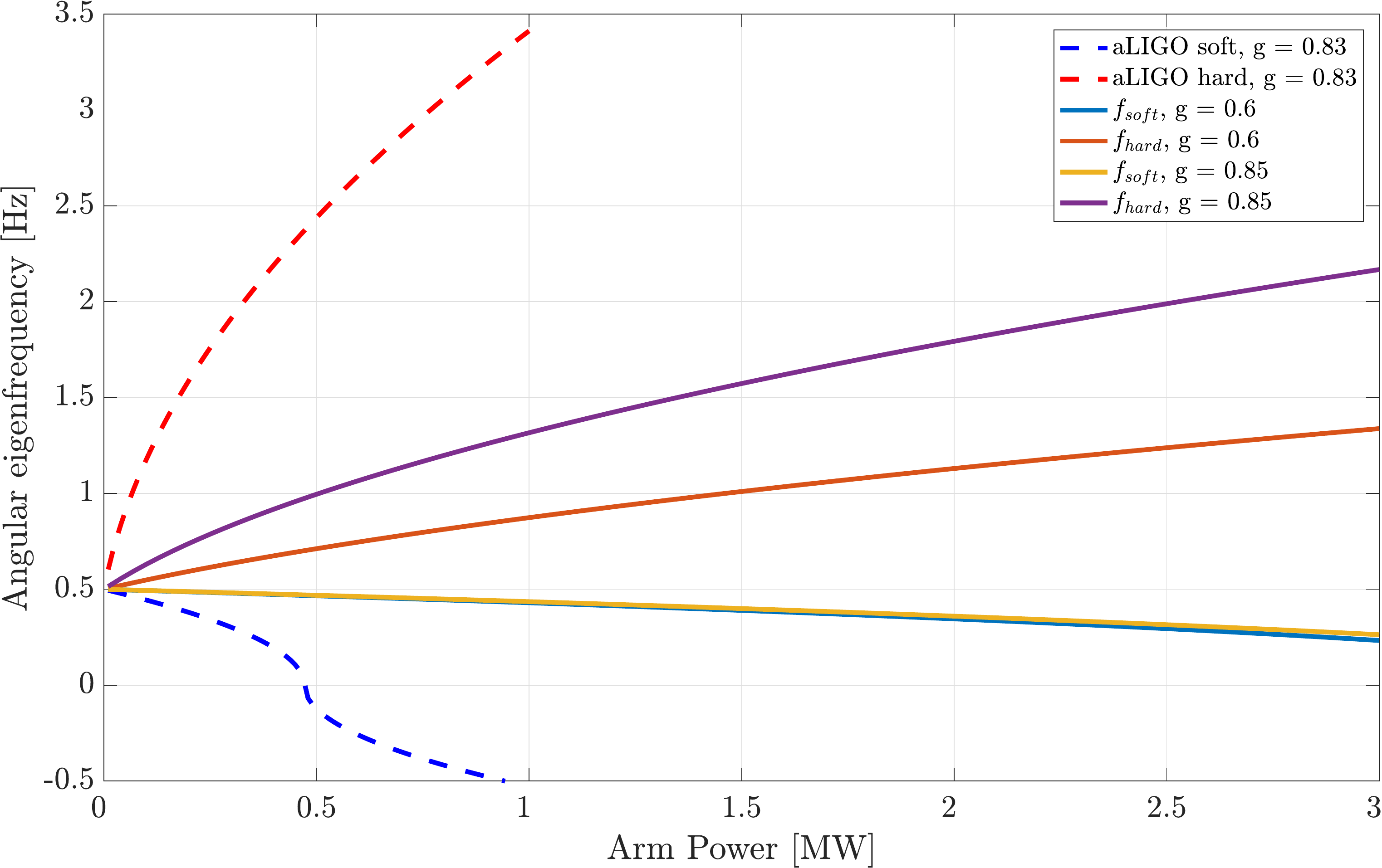}
  \caption[Angular Instability v. wavelength and g-factor]
          {Frequencies of the hard and soft modes vs.~arm power for various $g$-factors. The radii of curvature of the input and end test masses are set to  be the same in this simulation. The larger rotational moments of inertia for \voy{} remove the possibility of angular instabilities (a major controls problem in Advanced LIGO).}
  \label{fig:HardSoftModes}
\end{figure}

\Cref{fig:HardSoftModes} shows the eigenfrequencies of hard and 
soft modes for different power levels. 
Here the nominal laser wavelength of \SI{2000}{\nano\meter} for Voyager and 
\SI{1064}{\nano\meter} for Advanced LIGO is assumed. When the optical power is high 
enough, the soft mode becomes unstable. A robust feedback control loop 
should have enough bandwidth to suppress the instability. Simulations show 
that if the frequency of the unstable mode is 
$f_{\rm soft}$, then the bandwidth of the control loop needs to be 
$\sim 3 f_{\rm soft}$, and significant filtering of the 
sensing noise ($\sim60$\,dB) can be achieved at $\sim10 f_{\rm soft}$. 
Since the frequency of the soft mode is less than 1\,Hz for the Voyager 
design at \SI{3}{\mega\watt}, sensing noise from angular loops should not limit 
the sensitivity.

The hard/soft frequencies are functions only of the cavity $g$-factors, and not explicitly the laser wavelength. However, if the laser beam spot size on the mirrors is kept to a maximum value, $\omega_{max}$, due to clipping losses, then the cavity $g$-factor will be smaller for a longer wavelength.
Stated another way, if the beam spot size is maximized to reduce the thermal noise, the longer wavelength results in a more stable interferometer.

%% file: PI.tex
\label{s:PI}
The optical cavities and the interferometer mirrors have high quality factors, which allow for highly amplified resonances in the system. The accidental overlap of the resonances can lead to parametric instabilities (PIs)~\cite{Evans:2015}. As shown in Advanced LIGO, these are mostly mitigated by acoustic mode dampers (AMD) which are tuned masses bonded to piezo-electric transducers (PZTs) electrically connected to a dissipative element (resistor) \cite{Gras:2015}.  It has been found that one order of magnitude suppression
is easy to achieve.

Following the method of \cite{Gras:2015} we first compute the complex mechanical impedance of the mirrors. This model includes mechanical losses due to the coatings on the HR, anti-reflective (AR), and barrel surfaces. The baseline dimensions of the Voyager test mass have been used. The optical model considers the round trip optical gain, including scatter losses and clipping losses inside the arm cavities, as well as the optical transmissivities of the nominal \voy{} design.

There are still open questions regarding operation of AMDs in Voyager. Since Voyager is a cryogenic detector, the material selection for the AMDs must be reconsidered. It is well known that PZT performance is strongly temperature dependent. The properties of PZTs and bonding epoxies at \Tcryo{}\,K need to be examined.

\begin{figure}[ht]
  \centering
  \includegraphics[width=\columnwidth]{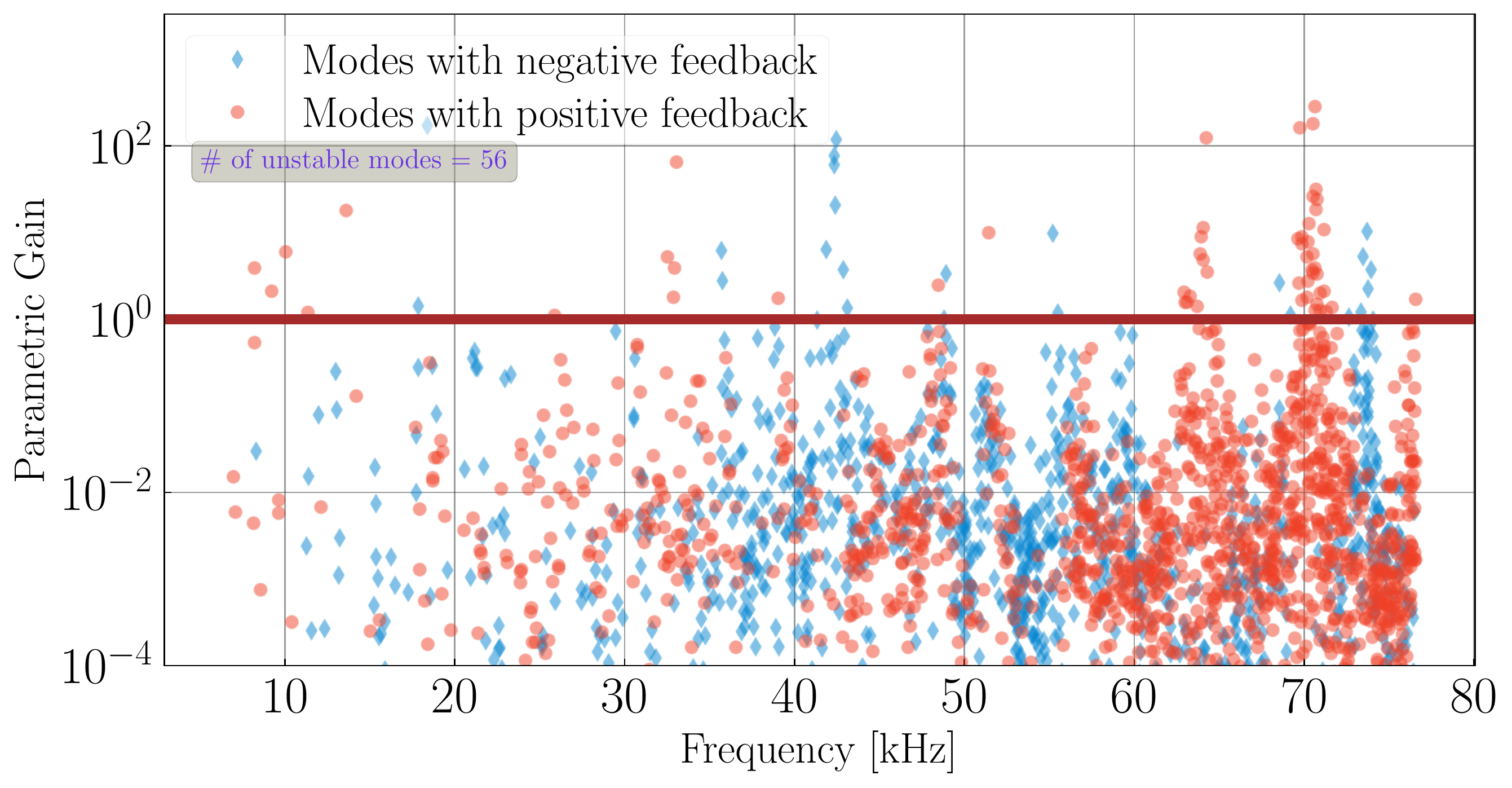}
  \caption[Parametric Instability gains]{Estimation of parametric gain
  for all of the opto-mechanical modes for LIGO Voyager. All of the
modes with positive gain (RED) greater than unity are considered to be
unstable. For this simulation, random perturbations have been added to the
RoC of the test mass optics.}
  \label{fig:PIgains}
\end{figure}

In \Cref{fig:PIgains}, we assume the nominal \SI{2000}{\nano\meter} operating wavelength and the concomitant test mass radii of curvature and cavity $g$-factors. Our stability analysis shows that there would be about 65 unstable modes without the application of AMDs --- significantly more than in Advanced LIGO. All of the unstable modes below 60\,kHz are weakly unstable. As the damping efficacy of the AMDs has a frequency dependence, we feel confident that AMDs can be designed for \voy{} which stabilize all of the modes without compromising the test mass thermal noise below 1\,kHz. Most likely, the \voy{} AMDs will use higher order mechanical resonances to damp the modes in a more frequency selective way.
There is very little change in the number or strength of the instabilities as a function of laser wavelength in the 1800\,--\,2100\,nm band
\footnote{Monte Carlo studies done of changing the Gouy phase of the power and signal recycling cavities, shows only a $\sim10$\% variation in the number of unstable modes.}
.

%% file: Quantum.tex
\label{s:Quantum}

\voy{}, like Advanced LIGO, will be limited by quantum noise in the majority of its  detection band, as illustrated in \Cref{fig:noise_comparison}. Extensive research has been carried out in the last decade to find solutions 
to reduce quantum noise in gravitational wave detectors.
The main approach relevant to \voy{} is squeezed light injection. Work has also been done on alternative interferometer optical topologies to recycled Michelson interferometers~\cite{Knyazev2018}, but these are not discussed here. Squeezing is a well-tested technology and was demonstrated in GEO600 and in the H1 LIGO interferometer~\cite{Gro2013, GEO:Squeezing, H1:Squeezing}, in preparation for use in the second generation detectors like Advanced LIGO and Advanced Virgo.

\begin{figure}[bthp]
  \centering
  \includegraphics[width=\columnwidth]{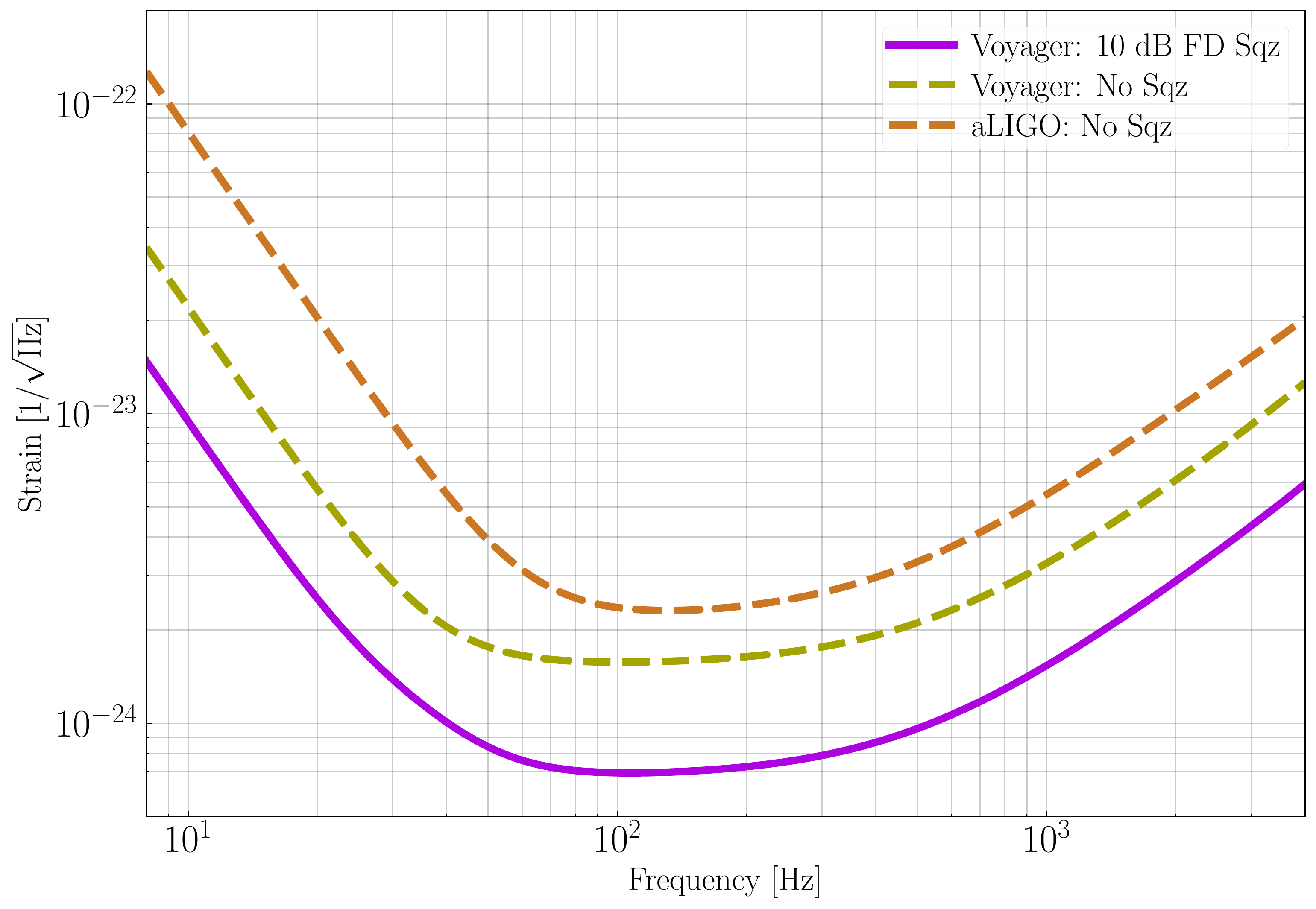}
  \caption{Quantum noise (QN) in \voy{} with 200\,kg test masses, \LaserPower{} of input power into the interferometer and +10\,dB of frequency-dependent injected squeezing. Also shown is the quantum noise with no squeezing. The region below 60\,Hz is dominated by radiation pressure noise. The region above 60\,Hz is dominated by shot noise.}
\label{fig:QN}
\end{figure}

Based on the analysis described in~\cite{Haixing:CC}, the \voy{}  
design relies on the injection of \SI{2}{\micro\meter} squeezed vacuum with a frequency dependent
squeezing quadrature~\cite{unruh1983quantum, KLMTV2001, PhysRevLett.116.041102} as a 
 solution to achieve a broadband reduction of quantum noise with 
respect to Advanced LIGO.  This is illustrated in \Cref{fig:QN} which shows frequency-dependent squeezing achieving a substantial improvement over the Advanced LIGO quantum noise floor. 
It can be seen that squeezing affords \voy{} a factor of two or more improvement in sensitivity across virtually the whole detection band
versus an unsqueezed \voy{}.

The design sensitivity curve shown in \Cref{fig:noise_comparison} and \Cref{fig:QN}
are obtained by injection of \SI{\GwincVal{Blue.Squeezer.AmplitudedB}}{\deci\bel} of 
squeezing at \GwincVal{Blue.Laser.Wavelength_nm}\,nm into a \GwincVal{Blue.Squeezer.FilterCavity.L}\,m long filter cavity with \GwincVal{Blue.Squeezer.FilterCavity.Lrt_ppm}\,ppm round-trip losses.
The loss between the squeezed light source and the interferometer is 
\pgfmathparse{int(round(\GwincVal{Blue.Squeezer.InjectionLoss}*100))}\pgfmathresult\%,
while the detection 
efficiency is 
\pgfmathparse{int(round(\GwincVal{Blue.Optics.PhotoDetectorEfficiency}*100))}\pgfmathresult\%,
yielding approximately \SI{7}{\deci\bel} of effective squeezing.

To ensure that the squeezing design will be feasible for \voy{}, we consider the state-of-the-art of three outstanding issues:

\begin{itemize}
\item squeezed vacuum generation at \SI{2000}{\nano\meter},
\item filter cavities for frequency-dependent squeezing in the audio band,
\item loss control for the squeezing system.
\end{itemize}


\subsection{Squeezed vacuum generation for 2000\,nm} 
\label{s:2umsqz}

By employing a 
coherent control scheme~\cite{Henning:PRL2006}, as typically done to produce squeezing at 
\SI{1064}{\nano\meter} in the audio frequency regime, high levels of squeezing down to 10\,Hz should, in 
principle, be obtainable at different wavelengths. Indeed, high levels of squeezing at \SI{1550}{\nano\meter} have already been demonstrated in the MHz 
regime (12.3 dB~\cite{Mehmet:2011je}) by pumping PPKTP at \SI{775}{\nano\meter}.  

\begin{figure}
  \centering
  \includegraphics[width=\columnwidth]{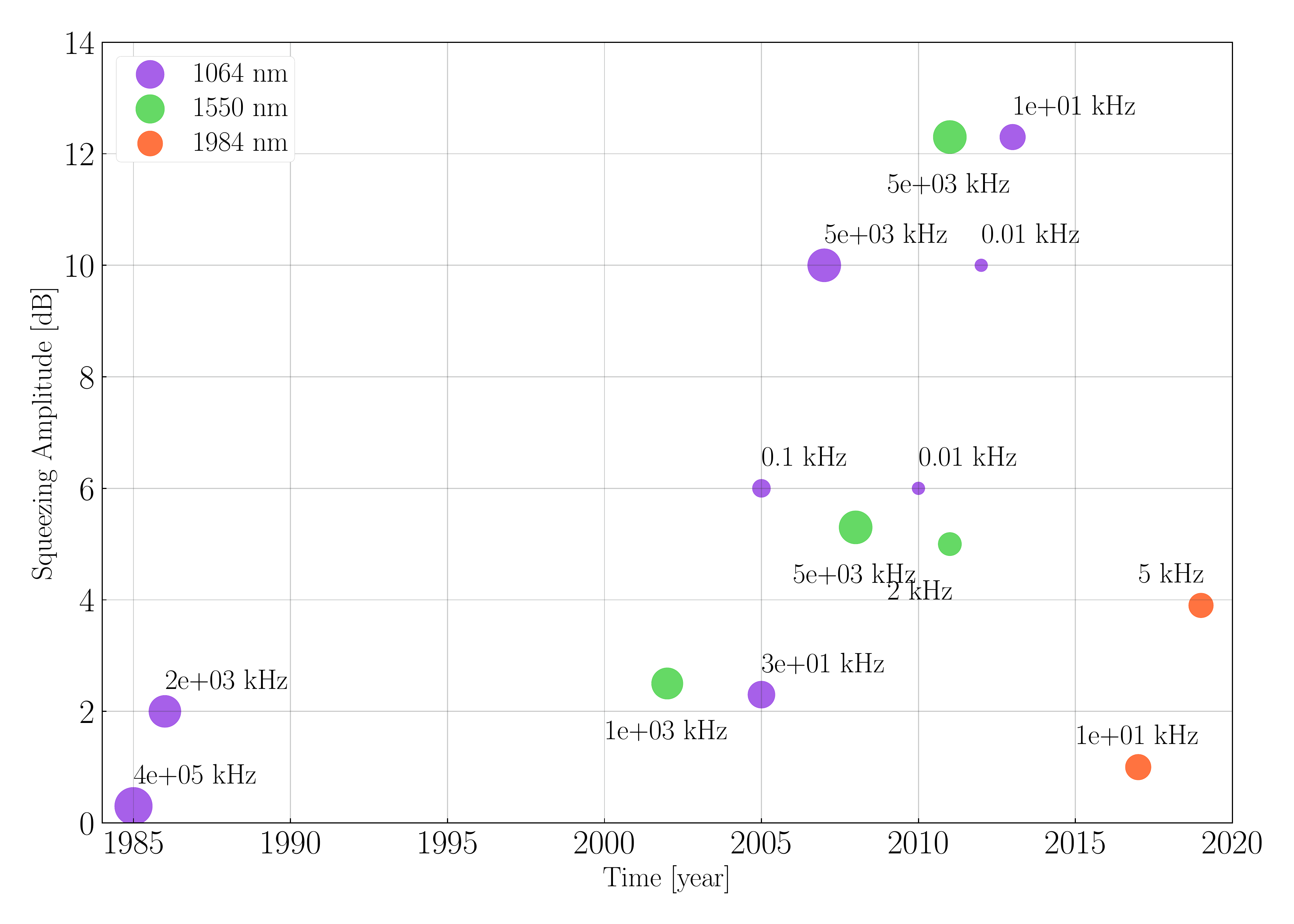}
  \caption[Squezing History]{Development of squeezed light sources over the years.
  The diameter of the circles is proportional to the log of the frequency at which
the squeezing was demonstrated (i.e. small circles are for low audio frequencies) \cite{Slusher:1985,Kimble:1986,Vah2008}}
\label{fig:SqueezeHist}
\end{figure}

At the moment, no new technical difficulties peculiar to \SI{2000}{\nano\meter} are anticipated and the best achieved squeezing to-date (at or near \SI{2000}{\nano\meter}) is  4\,dB in the 1\,--\,40\,kHz band, demonstrated with a laser source at \SI{1984}{\nano\meter}~\cite{Mansell:2018}. This is illustrated in \Cref{fig:SqueezeHist} along with the history of squeezed light generation at different wavelengths (provided for reference).

\subsection{Filter Cavities for Input Squeezing}
\label{sec:filter_cavities}

Quantum noise appears in two forms: shot noise and radiation pressure (back-action) noise.
Frequency-independent squeezed vacuum injection yields a reduction of high frequency quantum 
shot noise and a corresponding increase of low
frequency quantum radiation pressure noise. 
In this form, squeezed vacuum injection 
(as in~\cite{GEO:Squeezing, H1:Squeezing}) will not be suitable for \voy{}. 
However, squeezed vacuum can be manipulated to generate frequency-dependent squeezing
by rotating the squeezed field relative to the interferometer field in a frequency dependent way. 
This can be achieved by reflecting the squeezed beam from a high finesse, detuned 
{\it filter cavity} before injection into the interferometer~\cite{KLMTV2001}. 

Filter cavities and their properties have been extensively studied 
theoretically~\cite{Harms:2003wv, Kha2010, evans2013realistic}. The performance of a filter cavity 
can be characterized in terms of its intra-cavity losses per unit length. The lower the losses per 
unit length, the better the filter cavity is able to rotate the squeezing ellipse
without degrading it. 
Direct measurements report round-trip losses of 
10\,ppm (5\,ppm per bounce) for beam spot sizes in the 1\,--\,3\,mm range (corresponding to 
confocal lengths of 5\,--\,25\,m range), giving losses per unit length of 0.5\,ppm/m with a 
20\,m long filter cavity~\cite{Iso13}.

Frequency dependent squeezing at \SI{1064}{\nano\meter} has been experimentally demonstrated with rotation of the squeezing quadrature taking place around 1\,kHz and squeezing levels of 5.4\,dB and 2.6\,dB observed at high and low frequency, respectively \cite{PhysRevLett.116.041102}. Technical noises (optical loss, phase noise) have been recently calculated in order to estimate realistic performance of a filter cavity~\cite{Kwee:2017}.

The experimental characterization of the noise coupling mechanisms which limit the filter cavity performance is the next necessary milestone before validating this technology for application in gravitational wave detectors. The LIGO Scientific Collaboration has a program in place to achieve these goals for \SI{1064}{\nano\meter}.

A similar program needs to be established for \SI{2000}{\nano\meter}. A time scale of 3 years seems adequate to finalize a filter cavity design for \voy{}, informed by the outcome of the on-going effort for \SI{1064}{\nano\meter}.

\subsection{Loss Control: General}
In table top experiments, squeezing levels higher than 10\,dB have been measured~\cite{Vah2008, Stefszky:Balanced2012, Lisa:RPP:2018}.
However, the measured squeezing in GW detectors is strongly dependent on the total loss that the squeezed beam encounters in the path from the squeezed light source to the measurement photodetector. In practice, in existing gravitational wave detectors, reducing these optical losses below 20\% is non-trivial due to the large number of optical loss sources. For example, GEO600 reports~\cite{dooley2016geo} up to 4\,dB of detected squeezing, corresponding to 40\% total losses. \voy{} will have to contend with the same practical issues.

Every optical loss in the path from the squeezed light source to the final 
photo-detector contributes vacuum fluctuations that degrade the squeezed state. 
The list  of optical loss sources includes: squeezer optical parametric oscillator (OPO) internal losses, mode-mismatch, Faraday rotators and associated elements, signal- and arm-cavity losses, output mode cleaner (OMC) throughput and photodetector quantum efficiency (QE). To achieve 
\pgfmathparse{int(round(\GwincVal{Blue.Squeezer.InjectionLoss}*100))}\pgfmathresult\%  injection loss and 
\pgfmathparse{int(100- round(\GwincVal{Blue.Optics.PhotoDetectorEfficiency}*100))}\pgfmathresult\%, readout loss,
as currently assumed in the \voy{} baseline curve, all of these 
loss sources need to be of the order of 0.5\%\,--\,2\%.

Active mode-matching systems are currently in development for Advanced LIGO and we plan to continue development for \voy{}. 
Continued effort is required to develop low-loss small optics for the OMC, polarizing components, OPO and Faraday isolators.

\input{QuantumEfficiency}

\subsection{Conclusion}

The parts required for 10\,dB of audio-band frequency-dependent squeezing at \SI{2000}{\nano\meter} have yet to be demonstrated. Analogous demonstrations and the rate of technological development of squeezing over the last ten years, coupled with no fundamental reasons against, lead us to conclude that the \voy{} squeezing design is achievable.

%% file: QuantumEfficiency.tex
\subsection{Loss Control: Quantum efficiency}
\label{s:QE}
One of the most challenging loss considerations at \SI{2000}{\nano\meter} is the QE of photodiodes.
The QE of the photodetectors used at the GW signal extraction ports
must be {\textgreater}99\% with a goal of 99.5\%.
Additionally, the high-QE photodiodes will need to remain linear and low-noise  with approximately \SI{10}{\milli\watt} of optical power incident on them.
Several options are available for detectors: extended InGaAs, mercury cadmium telluride (MCT or HgCdTe), and InAsSb, and these are discussed below. At this time, none of these options meets the high-QE, linearity, and low noise requirements, and significant development will be required on all these technologies to achieve better than 99\% QE while simultaneously coping with the large amount of incident optical power.

\subsubsection{Extended InGaAs photodetectors}

Current extended InGaAs photodetectors typically have low QE ($\sim$75\%) around \SI{2000}{\nano\meter}, although
Laser Components Inc.~has a series of photodiodes that have QE up to 87\%~\cite{LaserComponentsExtendedInGaAs}.

Extended InGaAs photodiodes achieve a broader spectral response by varying the relative amounts of InAs and GaAs in the semiconductor alloy to 
increase the cut-off wavelength of the photodetector. Unfortunately,  
this leads to lattice spacing mismatch within the material that, in turn, results in significantly increased $1/f$ noise and, indirectly, lower QE. It is an active area of research to determine if QE can be increased in extended InGaAs without introducing catastrophic levels of low-frequency dark noise.

\subsubsection{HgCdTe (MCT) photodetectors}

Mercury cadmium telluride (MCT or HgCdTe) detectors are commonly used for infrared astronomy. They have a strong response in the mid-IR from 1.5\,\si{\micro\meter}, with cut-off wavelengths of 2.5\,\si{\micro\meter}, 5\,\si{\micro\meter} or longer, depending on the construction. MCT detectors with an broadband AR coating have measured QE of approximately 94\% \cite{Teledyne:2017}.

As most MCT detectors are p-n junction based, they rely on diffusion of electrons and holes across the active region which is a (relatively) slow process and can lead to recombination of holes and electrons before they reach the junctions. Recombination results in an effective loss of QE as those charge carriers are ultimately not converted into photocurrent. MCT photodiodes could be promising in configurations other than p-n junction.

\subsubsection{InAsSb photodetectors}

InAsSb detectors have matured in the past two decades \cite{Martyniuk:2014}. Traditionally low QE and high noise (when used at room temperature), InAsSb performance has improved in recent years by exploiting  different junction architectures \cite{Kilpstein:2011,Soibel2011, Steenbergen:2011}. They are currently a promising candidate for further consideration.

%% file: Suspensions.tex
\label{s:Suspensions}
\subsection{Introduction}

The \voy{} suspension system will have much in common with the
Advanced LIGO suspension~\cite{Aston:2012}. The basic quadruple
pendulum design will be used.
The upper two masses and their suspensions will be made from steel,
and the lower two masses and suspension elements between them are made from a single material
(silica in Advanced LIGO and silicon in \voy{}).
Hydroxide-catalysis bonding~\cite{Rowan:1998,vanVeggel:2009} or
optical contacting will be used to assemble the final monolithic stage.
The three-stage seismic isolation system
used in Advanced LIGO~\cite{Wen:2014hm, Matichard:2015hb}
will be reused for \voy{}, with minor engineering modifications
to accommodate the heavier payload.

There are, however, two major differences between the Advanced LIGO
suspensions and those of \voy{}:
(i) The silica cylindrical fiber final stage suspension will be
replaced with silicon ribbons,
and (ii) silicon cantilever blade springs for vertical isolation
will be added to the final stage.

\begin{figure}[b]
  \centering
  \includegraphics[width=0.5\columnwidth]{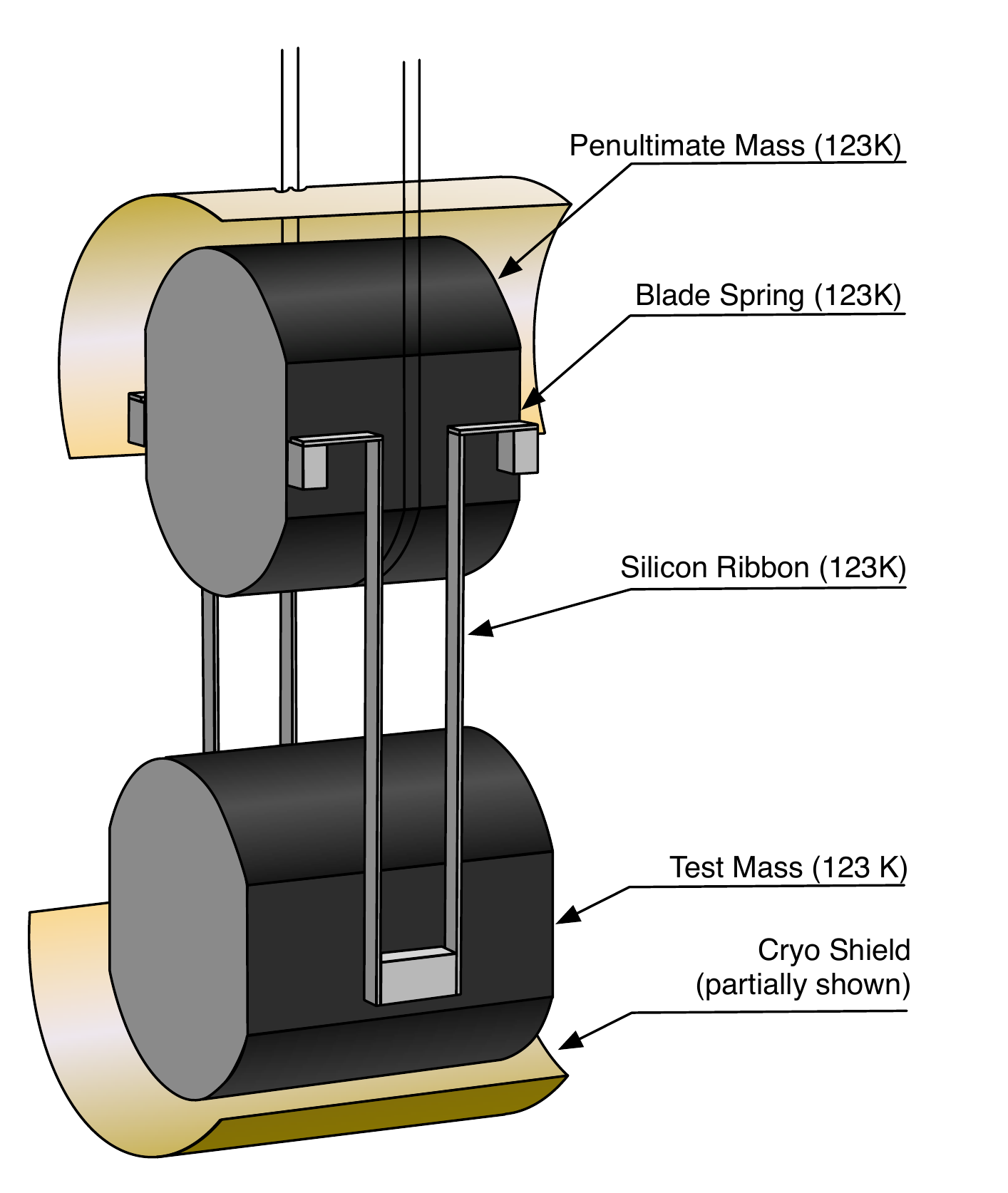}
  \caption{Conceptual model of the \voy{} silicon monolithic
  suspension. The plates surrounding the masses represent a cut-away
  view of the thermal shields.
  }
  \label{fig:suspension}
\end{figure}

\subsection{Suspension design}
The lower two masses of
the \voy{} suspension will be cooled radiatively (\Cref{fig:suspension}).
The silicon test mass will be suspended by four silicon ribbons, via silicon vertical-spring blades attached to the silicon penultimate mass.
This section between the test mass and the penultimate mass is
conductively cooled by the cold masses.
The cold section and the other upper masses are suspended with steel wires from the upper stages.

The mass distribution and suspension lengths have been
designed to minimize the quadrature sum of the modeled seismic
noise and suspension thermal noise at \SI{12}{\hertz}, as described in
\Cref{tab:suspension_mass_distribution}.
The current seismic platform is able to support a payload of up to \SI{1150}{\kilo\gram}.
In our design, 
\SusTotalMass~\SI{}{\kilo\gram} was assigned to the main suspension chain, reserving \SI{630}{\kilo\gram} for the reaction chain, the cage
structure, and balancing mass.
The total length of the main suspension chain from the top
suspension point to the optical height of the test mass remains the
same as Advanced LIGO.

\begin{figure}
  \centering
  \includegraphics[width=\columnwidth]{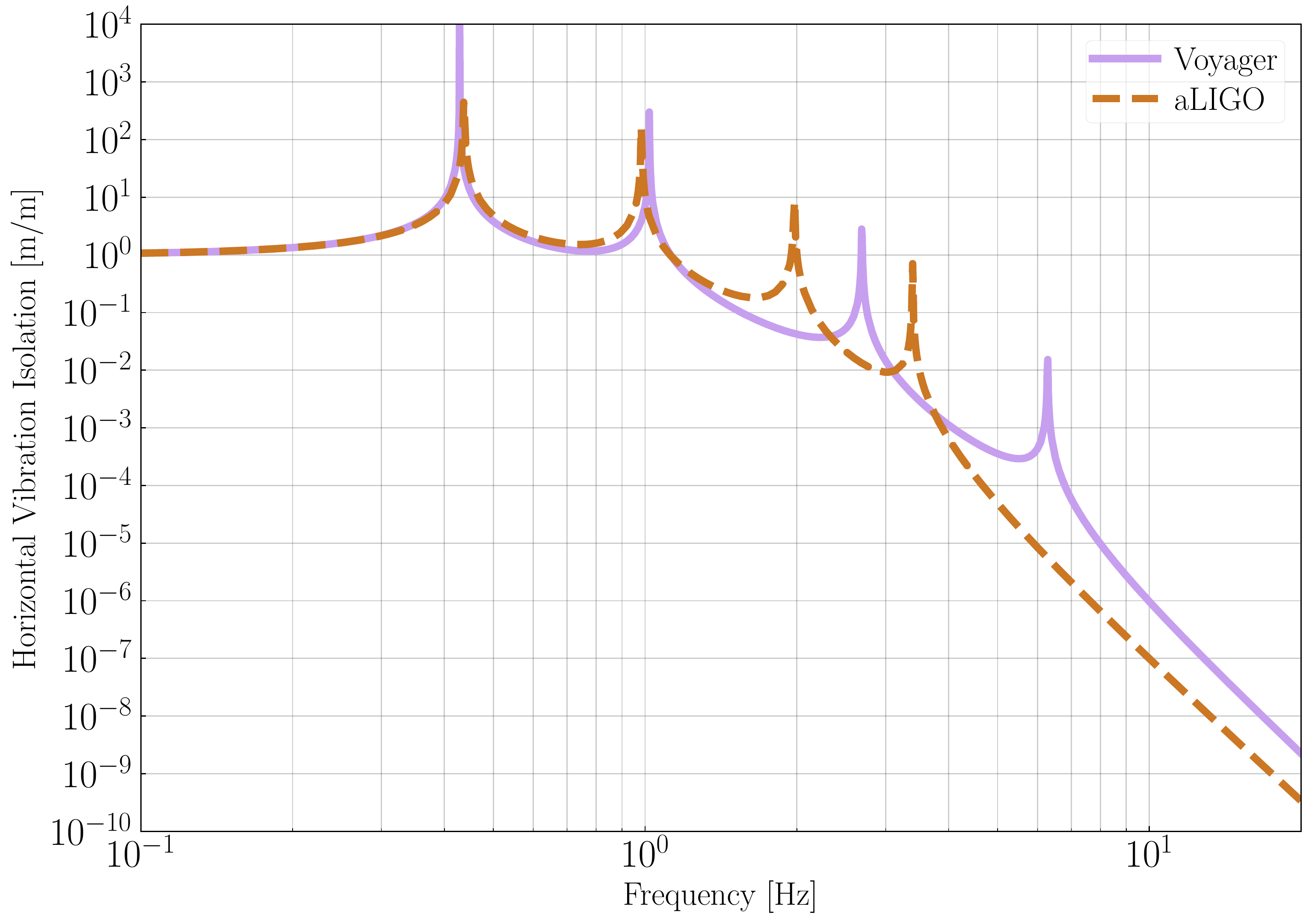}
  \caption[Vibration isolation of the suspension]
  {Horizontal vibration transmission of the quadruple suspensions in Advanced LIGO and \voy{}.}
  \label{fig:suspension_isolation}
\end{figure}

The resulting overall isolation of the suspension is shown in
\Cref{fig:suspension_isolation}.
For a given total length of the suspension chain, the best vibration
isolation above the pendulum resonant frequencies is realized with
equal length stages~\cite{T1300786}.
However, we have chosen to make the bottom two stages as long as possible, so as to reduce the thermal noise from the penultimate stage (cf.~\Cref{sec:sus-thermal-noise}).
To maintain the $\sim$\SI{10}{\hertz} seismic wall, the noise
of the seismic platform can be improved through lower noise seismometers for the feedback control.

\begin{table}[b]
\centering
\begin{tabular}{lccccc} \toprule 
&\multicolumn{2}{c}{aLIGO} &\hspace*{10pt}& \multicolumn{2}{c}{\voy{}} \\ 
Parameters & mass (\SI{}{\kilo\gram}) & length (\SI{}{\meter}) && mass (\SI{}{\kilo\gram}) & length (\SI{}{\meter}) \\ \midrule 
Payload total    & 124 & 1.642 && \SusTotalMass & 1.642 \\
Top mass         &  22 & 0.422 && \SusMassTop   & \SusLengthTop \\
Second mass      &  22 & 0.277 && \SusMassUIM   & \SusLengthUIM \\
Penultimate mass &  40 & 0.341 && \SusMassPUM   & \SusLengthPUM \\
Final mass       &  40 & 0.602 && \SusMassTM    & \SusLengthTM \\ \bottomrule
\end{tabular}
\caption[Summary of the suspension parameters]
{Summary of the suspension parameters for the quadruple suspensions
 for Advanced LIGO and \voy{}. Here the length of each stage refers to the
 wire (ribbon) length between that stage and the one above it. The
 total length refers to the total length of the suspension chain from
 the top suspension point to the optic
 center.}  \label{tab:suspension_mass_distribution}
\end{table}

\subsection{Fabrication of a monolithic silicon final stage}
The final stage suspension will use silicon ribbons to suspend the
silicon test mass from silicon vertical-spring blades bonded to the
silicon penultimate mass.
Crystalline silicon is the preferred material for the suspension, considering the thermal noise and the material matching with the mirror.
Silica fibers like those of the second generation detectors are not suitable, because of their increased mechanical dissipation at low temperature~\cite{ANDERSON:1955cb, Fine:2004hr, Marx:2004dr, McSkimin:2004dh}.
Not all of the engineering design of the monolithic stage has been determined. However, as discussed in \Cref{sec:sus-thermal-noise}, the thermal noise in this stage does not limit the sensitivity of the interferometer. It leaves plenty of room to relax the design requirements regarding the thermal noise to make the construction of this stage feasible. 

\subsubsection{Production of silicon ribbons}

Silicon ribbons can be manufactured by cutting and etching a long silicon boule or a large wafer~\cite{Cumming:2014iz}.
In the \voy{} design, the ribbons have a width of \SI{10}{\milli\meter}
and a thickness of \SI{0.5}{\milli\meter}.
Since the test mass is cooled by
radiation, the ribbon dimensions are determined purely by the tensile
strength of silicon, and heat conduction is irrelevant.  A
review of silicon's tensile strength can be found
in~\cite{Cumming:2014iz}. There the measured tensile strengths range
from \SI{200}{\mega\pascal} to \SI{8.8}{\giga\pascal}, depending on
the dimensions of the ribbons, and the importance of the surface and
edge quality is emphasized.  Recent results under various surface
treatments are found in~\cite{Cumming:2014iz}
and~\cite{Birney:2017dt}, and there the average tensile strengths are
distributed from \SI{100}{\mega\pascal} to \SI{400}{\mega\pascal}.
We have assumed a tensile strength of \SI{100}{\mega\pascal} to provide
a safety factor, although stronger and thinner ribbons should become
possible as the development progresses.

\subsubsection{Hydroxide-catalysis bonding of the final stage}
Instead of the laser welding used for fused silica suspensions,
hydroxide-catalysis bonding (HCB) can be used for the assembly of the
\voy{} suspensions. HCB of oxide
materials~\cite{Rowan:1998} was used in Gravity Probe
B~\cite{Buchman:1996fl}, and also to bond some glass parts to
aLIGO test masses~\cite{Aston:2012}.
The same technique has been demonstrated to work on silicon~\cite{vanVeggel:2009}.
The upper limit of the mechanical loss associated with the bonded silicon was reported to be $\phi < (5 \pm 2) \times 10^{-3}$~\cite{Prokhorov:2017jd}.
The effect of this mechanical loss is not included in the thermal noise calculations here, and will have to be calculated by FEA. As with Advanced LIGO, we expect to only estimate the thermal noise using FEA and mechanical Q measurements, since the direct measurement of the suspension's thermal noise is challenging even with the extremely low displacement noise of a gravitational wave interferometer.

\subsubsection{Vertical suspension isolation}
Vertical-spring blades are designed to lower the vertical resonant
frequency and reduce the amount of vertical seismic and thermal
noise coupling to the horizontal motion of the mirror. The bottom
stage vertical springs will be made of silicon, for cryogenic
compatibility, and they will be HCB-bonded to the penultimate mass and
to the ribbon, as conceptually shown in \Cref{fig:suspension}
(cf.~the sapphire blades for KAGRA~\cite{2016JPhCS.716a2017K}).

As with silicon ribbons, the dimensions of the silicon vertical-spring blades  will be determined by the breaking stress of silicon.
If a breaking strength of \SI{100}{\mega\pascal} is assumed,
a \SI{400}{\milli\meter}-length \SI{80}{\milli\meter}-wide triangular
blade with a thickness of \SI{12}{\milli\meter} can
sustain \SI{50}{\kilo\gram} of load.
This blade would have a vertical spring constant of \SI{6.5e4}{\newton / \meter}, and yields a rather high resonant frequency of \SI{5.7}{\hertz}. Further surface treatments of the blades should should allow us to
increase the breaking strength and lower the vertical frequency.

\begin{figure}[h]
  \centering
  \includegraphics[width=\columnwidth]{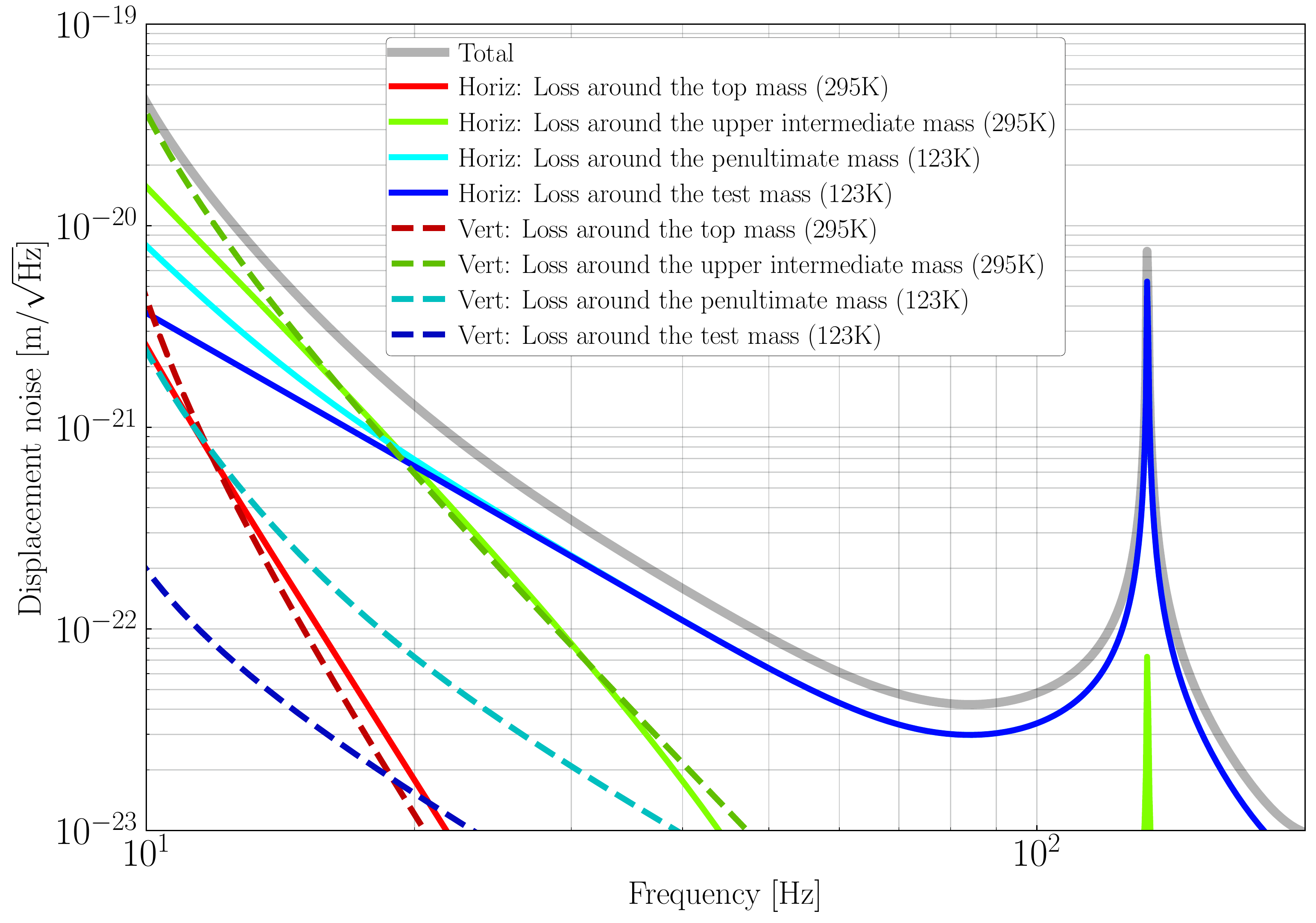}
  \caption[Suspension Thermal Noise]
  {Estimated total suspension thermal noise (for a single test mass), and the contribution from losses localized at each mass.}
  \label{fig:susnoise}
\end{figure}

\subsection{Suspension thermal noise}
\label{sec:sus-thermal-noise}
\Cref{fig:susnoise} shows the current thermal noise estimate for a single test mass suspension, along with a breakdown of the losses associated with each suspension stage.
This rough thermal noise model includes the bulk mechanical loss of the silicon and steel for the ribbons, wires, and blades, thermoelastic noise, and the surface loss effect of silicon.
It does not include more detailed effects such as the mechanical loss associated with bonding, the shape factor of the ribbons and blades, couplings between the mechanical modes, etc.
An increase in the suspension thermal noise relative to our rough estimates will result in a small increase in the overall \voy{} noise at $\sim10$\,Hz, and a negligible impact on astrophysical metrics~\cite{VoyScienceCase}.

The total suspension thermal noise below \SI{20}{\hertz} is dominated by the
penultimate stage wire, particularly its upper end which is attached to
the upper intermediate mass.
This is due to the high mechanical loss ($\phi = \SI{2e-4}{}$) of the steel wire, and the warm temperature of the mass.
This noise cannot be mitigated by, for example, changing the wire material to fused silica or silicon. The penultimate stage has different temperatures for the upper and lower joints, and these alternate materials would suffer from excessive thermal noise at either the upper or the lower joint.
To filter the noise from the penultimate stage, the lengths of the final two stages were made as long as possible (leaving \SI{0.15}{\meter} as the minimum
length of the upper stages). The noise of the final stage dominates above \SI{20}{\hertz}, but is very low due to the low mechanical loss and negligible thermal expansion of silicon at \Tcryo{}\,K.

Another notable feature is the violin mode seen around \SI{120}{\hertz}, as determined by the size of the silicon ribbons. Further improvement of the tensile strength and optimization of the ribbon size is desirable, to move the violin modes higher.

%% file: Lasers.tex
\label{s:Lasers}

\voy{} will operate at a laser wavelength in the range between \SI{1900}{\nano\meter} and \SI{2100}{\nano\meter} (see \Cref{s:Wavelength} for a full discussion of the wavelength choice).  The final wavelength selection will depend both on other \voy{} subsystem requirements and also on the development of near-IR and mid-IR lasers over the next several years. For simplicity in the following text, we refer to this wavelength range as \SI{2}{\micro\meter}. 

In this section, we discuss the requirements and candidate technologies for the \voy{} pre-stabilized laser.
In considering those candidates, we will 
review the current state-of-the-art laser technology around \SI{2}{\micro\meter}. Commercial development of lasers within this wavelength range is of growing interest and is largely driven by remote sensing applications (e.g. spectroscopy of different gas species, particularly atmospheric CO2 and water \cite{Scholle2010}). Although promising free space and fiber lasers and amplifiers candidates exist, there is currently no commercial laser that meets all of the design specifications. This was also true when Initial and Advanced LIGO were in the same stages of development that \voy{} is today and is not concerning: we expect  mid-IR laser development to follow a similar trajectory to the one seen for \SI{1064}{\nano\meter} lasers in Advanced LIGO.

\subsection{Laser requirements}

The requirements for the \voy{} laser are summarized in \Cref{tab:laser_requirements}.

\begin{table}[h]
    \begin{center}
    \begin{tabular}{lcc}\\ \toprule
         {Type} & {Requirement}  & {Comment}   \\ \midrule
         Wavelength & \SI{1900}{\nano\meter}$\--$\SI{2100}{\nano\meter} & Single-frequency \\ 
         Power &  \LaserPowerPSL{} & CW operation \\
         Polarization* & horizontal,  $>$100:1 ratio \cite{T050036} & \\
         Spatial mode* & $\ge$97\% TEM00  \cite{T050036} & \\
         HOM content* & $<$ 3\%  \cite{T050036} & \\
         Intensity noise (RIN)* & $\le 10^{-6}$ Hz$^{-1/2}$  & \SI{10}{\hertz} $\le f \le$ \SI{5}{\kilo\hertz} \cite{Heurs:04} \\
          & $\le 2\times 10^{-7}$ Hz$^{-1/2}$  & \SI{10}{\kilo\hertz} $\le f \le$ \SI{10}{\mega\hertz} \cite{T050036} \\
          & $\approx$ shot-noise limited &  $f \ge$ \SI{10}{\mega\hertz} \cite{T050036} \\ 
         Freq.~noise (free-running)* & $\le \left(10\, \mathrm{kHz}/f\right)$ Hz$^{-1/2}$ &  \SI{1}{\hertz} $\le f \le$ \SI{5}{\kilo\hertz} \cite{Kwee:12}\\
         Freq.~actuation bandwidth* &  \SI{200}{\kilo\hertz} \cite{Hall2017} & \\
         Operation & stable 365/24/7  & no maintenance required\\
         Lifetime & 10+ years &  continuous operation\\
         \bottomrule
    \end{tabular}
    \caption{Provisional list of laser requirements. Those requirements marked with asterisks (*) are based on equivalent requirements or performance for the Advanced LIGO laser. Although linewidth is a popular specification, we specifically do not use it for characterizing frequency noise requirements.}
    \label{tab:laser_requirements}
        \end{center}

\end{table}

\subsubsection{Power}

The laser beam  must deliver approximately \LaserPowerApprox{} of stabilized single frequency power at \SI{2}{\micro\meter}  to the power-recycling mirror (PRM), as illustrated in \Cref{fig:IFO_schematic}.  Given that the total transmission of the input optics between the laser and PRM is approximately 70\%, the required output of the laser is approximately \LaserPowerPSL{}.

The current laser design is broadly similar to the existing Advanced LIGO laser \cite{Kwee:12}, based around three stages of increasing power:
\begin{itemize}
    \item Low power stage: requires a low intensity and phase noise, single-frequency, linearly polarized, CW, \SI{2}{\micro\meter}  master oscillator laser with output power of approximately \SI{1}{\watt} and good beam quality with ${\rm TEM}_{00}$ mode content preferably {\textgreater}97\%.
    \item Medium power stage: the laser enters a medium power second stage in which it amplified to  \SI{35}{\watt}. 
    \item High power stage: the last stage of the laser system amplifies the output to  \LaserPowerPSL{}.
\end{itemize}
\noindent The medium and high power stages must maintain the same low noise characteristics and good beam profile as the master oscillator.

\subsubsection{Remaining requirements}

Requirements for the higher-order mode (HOM) content and frequency and intensity noises are to be derived using a closed-loop, higher spatial order, opto-electronic feedback model of \voy{} that includes realistic assumptions about absorption, thermal lensing and compensation based upon experiences with Advanced LIGO. At this time this model is still being developed. We use the Advanced LIGO laser requirements as a guide for upper limits to these requirements. These are listed in \Cref{tab:laser_requirements}. These requirements are expected to be refined as realistic models of \voy{} are developed.

The \voy{} laser is expected to require at least a similar free-running frequency noise as Advanced LIGO. The frequency actuation bandwidth of the entire laser system  must be approximately \SI{200}{\kilo\hertz} \cite{Hall2017} in order to be able to sufficiently stabilize  frequency fluctuations to the low-noise reference of the interferometer itself. It is not necessary for each stage (master oscillator, medium power stage and high power stage) to be individually  capable of providing this bandwidth, provided that there is one or more component in the system that can.

The relative intensity noise (RIN) requirements, extending into the RF, are based on the intensity noise of the PSL in Advanced LIGO \cite{T050036}. Once integrated into \voy{}, the laser intensity noise will also be suppressed to a similar level to Advanced LIGO.

The laser is expected to run continuously without requiring major maintenance for the lifetime of the \voy{} project, at least 10 years.

\subsection{Laser candidate technologies and examples}

There are two rare-earth dopants suitable for direct lasing at \SI{2}{\micro\meter}: thulium and holmium, which can provide optical amplification in the \SI{1900}{\nano\meter}--\SI{2040}{\nano\meter} \cite{ThFiberLaserReview}, and  \SI{2040}{\nano\meter}--\SI{2170}{\nano\meter} \cite{HoLasersReview2014} bands, respectively.

Two basic laser architectures are available: optical fiber \cite{Hemming:15} and free-space \cite{Ganija:16} lasers. These architectures are not necessarily incompatible and the final system may contain a low-power free-space master oscillator (MO) followed by some combination of power oscillators and fiber amplifiers.

The following subsections detail examples of thulium and holmium lasers that are expected to meet the majority of the requirements for the \voy{} laser.

\subsubsection{Single-frequency, low-noise source}
\label{s:SFlaser}
The full laser system begins with a master oscillator stage that is a low-noise, single-frequency source.

Fiber laser master oscillators use short lengths of doped silica fiber with spectrally-matched distributed-Bragg-reflector (DBR) fiber gratings \cite{Fu:17} that are spliced onto each end. The gratings are fabricated to suit the required lasing wavelength of the Tm or Ho dopant, and thus a broad range of wavelengths are possible and modifying the wavelength of an otherwise suitable MO to satisfy other interferometer requirements should be possible. Achieving a stable, narrow linewidth MO will require careful thermal and vibration isolation of the fiber from the environment however.

Given that different wavelengths can be achieved, we must next consider the frequency noise. Determining if a commercial laser meets the frequency noise requirements from specifications alone is typically not possible as these lasers usually quote linewidth rather than frequency noise in their specifications. 
The single-frequency \SI{10}{\watt} Q-Peak Firebow CW10-500, which has a linewidth of $<$\SI{1}{\mega\hertz} \cite{Firebow:2018}, is a possible candidate, but the stability and frequency noise would need to be measured to verify compatibility with \voy{} requirements.

Alternatively, the free-space single-frequency non-planar ring oscillator (NPRO) architecture has been demonstrated to have low frequency noise, for example at \SI{1064}{\nano\meter} \cite{Willke:00}. 
This architecture uses a crystalline gain medium and thus only a few wavelengths are possible: a \SI{400}{\milli\watt} Tm:YAG NPRO at \SI{2013}{\nano\meter} \cite{Lin2009} and a \SI{7.3}{\watt}  NPRO at \SI{2090}{\nano\meter} \cite{Yao:08} have been reported but frequency noise spectra were not available at the time of writing.  Additionally, lasing of cryogenic Tm:YAG at \SI{1880}{\nano\meter} has been demonstrated \cite{Johnson1965} and is expected to be suitable for use in an NPRO.

Marginally outside the \SI{1900}{\nano\meter}$\--$\SI{2100}{\nano\meter} range is a \SI{2128}{\nano\meter} laser. Lasers at this wavelength do not yet meet all the requirements for \voy{} but could leverage existing \SI{1064}{\nano\meter} components for frequency stabilization (when doubled and locked to the existing aLIGO lasers).

The existence of single-frequency, narrow linewidth, \SI{2}{\micro\meter} lasers is encouraging but more work needs to be done to determine if the frequency noise of these lasers meets  the requirements of \voy{}.

\subsubsection{High power}
\label{s:HPlaser}

High power lasers that could serve as amplifiers have been demonstrated. For example, a multi-mode, CW, Ho-doped fiber laser at \SI{2100}{\nano\meter} has been demonstrated with  power up to \SI{400}{\watt} \cite{Hemming:13, Simakov:14}. 
Reviews of recent work in fiber lasers are provided by Hemming \cite{HoLasersReview2014} and Fu \cite{Fu:17}.

High power free-space CW oscillators have also been demonstrated, including a \SI{200}{\watt} Tm:YAG laser [ref] and a \SI{65}{\watt} cryogenic Ho:YAG laser that produced a \SI{100}{\watt} output at \SI{2097}{\nano\meter} with good beam quality \cite{Ganija:17}.

The closest example of an existing high-power low-noise laser is the \SI{600}{\watt}, \SI{2040}{\nano\meter} single-frequency single-mode thulium fiber laser demonstrated by Goodno~et~al.~\cite{Goodno:09} that amplifies a \SI{5}{\mega\hertz} linewidth distributed feedback laser diode from \SI{3}{\milli\watt} to greater than \SI{600}{\watt} and maintains the low linewidth of the source. For this laser, stimulated Brillouin scattering (SBS) was demonstrated to be  negligible below \SI{250}{\watt} output power and, as such, is not expected to be an issue for the \voy{} laser. Some early indications from high-power laser work at \SI{2}{\micro\meter} suggest that there may be power-dependent excess relative intensity noise at radio frequencies (RF) and this remains an active area of investigation.

\subsection{Summary of laser prospects}

Most of the constituent requirements of a pre-stabilized laser around \SI{2}{\micro\meter}  (master oscillator, intermediate amplification and high power stages) have  been demonstrated at or near this wavelength. Special emphasis needs to be placed upon acquiring frequency and intensity noise measurements on low-noise master oscillators soon.

It is clear that full confirmation of a \SI{2}{\micro\meter} laser source with sufficiently low frequency and intensity noise is still to be performed. We are confident that this can be achieved as the requirements for the \voy{} laser are not substantially beyond specifications already demonstrated at other wavelengths \cite{Kwee:12}. Subsequent development and engineering work are still needed to integrate all these parts into a single system, however no fundamental reasons preclude the production of such a system.

%% file: configurations.tex
\label{s:configs}

\begin{figure}[t]
  \centering
  \includegraphics[width=\columnwidth]{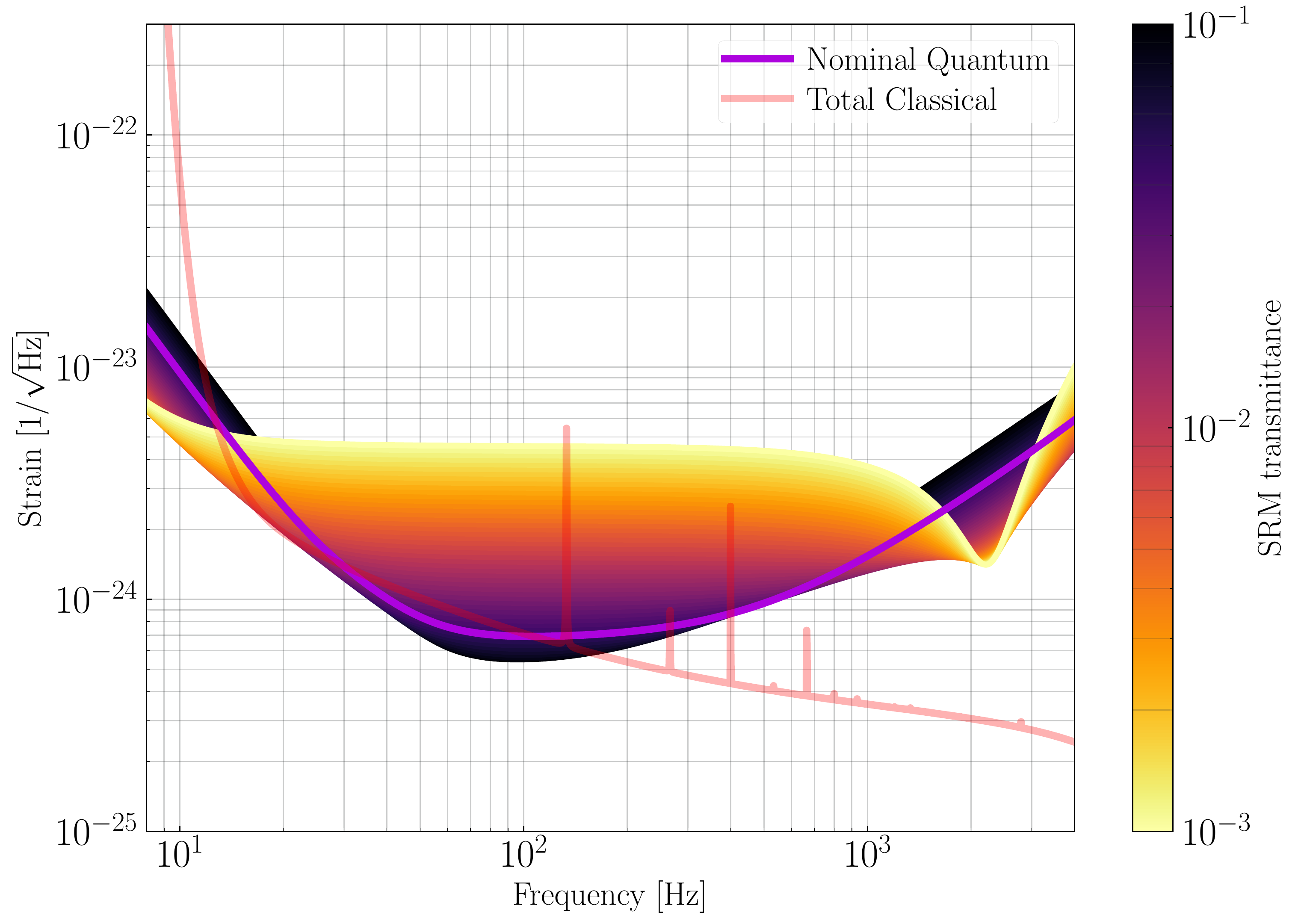}  
  \caption[Voyager Noise Curve vs SRM transmittance]
  {Modifying the \voy{} noise curve by replacing the SRM is a way to optimize the sensitivity to sources in the kHz band, such as binary neutron stars post-merger.}
  \label{fig:srmsweep}
\end{figure}

The nominal \voy{} design is optimized for broadband operation by balancing quantum and classical noise sources, so as to maximize the number of detections of binary neutron star systems. 
Varying the signal recycling mirror (SRM) reflectivity will produce a different quantum noise floor.
In this way, we can optimize for high frequency or low frequency operation, as is illustrated in Figure \ref{fig:srmsweep}.

The nominal SRM transmittance, $T_{\mathrm{SRM}}$, is  \SI[round-mode=figures,round-precision=2,scientific-notation=true]{\GwincVal{Blue.Optics.SRM.Transmittance}}{}. In Figure \ref{fig:srmsweep}, the interferometer has also been optimized for a range of transmittances between $1\times10^{-3}$ and 0.1. 
An SRM transmittance of 0.1 reduces the quantum noise in the band around \SI{30}{\hertz}$\--$\SI{300}{\hertz}. However, the quantum noise becomes lower than the classical noise in that band. As such, there will be limited overall sensitivity improvement without a significant reduction of classical (coating thermal) noise.

Conversely, if we use an SRM transmittance less than \SI[round-mode=figures,round-precision=2,scientific-notation=true]{\GwincVal{Blue.Optics.SRM.Transmittance}}{}, we see improved sensitivity above \SI{800}{\hertz}, at the expense of broadband sensitivity. Such a configuration would be useful, for example, for exploring the neutron-star equation of state. The dip in noise around \SI{2300}{\hertz} comes about because of the coupled cavity resonance between the arm and the signal recycling cavities~\cite{Martynov2018}.

%% file: conclusion.tex
\clearpage
\section{Conclusion}
\markboth{}{}
\label{s:conclusion}
We have described \voy{}, a design concept for the next generation of ground based gravitational wave detector. The design takes advantage of large silicon mirrors, operated at high optical power and cryogenic temperatures, with quantum assisted metrology.

This instrument will extract the full potential of the existing LIGO facilities. Nearly all of the existing infrastructure (including the complex vibration isolation systems) will be re-used, greatly reducing the cost and complexity of the upgrade.

Much of the R\&D required for \voy{} has been ongoing for several years to support the cryogenic KAGRA and Einstein Telescope designs, and will also be applicable to the Cosmic Explorer design~\cite{CosmicExplorerarXiv}.

We anticipate that \voy{} will open the next chapter of major discoveries in gravitational wave astronomy~\cite{VoyScienceCase}. The upgraded detectors will find thousands of binary neutron stars, and detect stellar-mass binary black holes from throughout the cosmological era in which such mergers are believed to have taken place.  The nearest sources will be detected with unprecedented clarity, providing highly sensitive probes of the behavior of ultra-dense matter and the nature of gravity itself.

 This work was supported in part by the National Science Foundation under the LIGO cooperative agreement PHY-0757058. This paper has been assigned LIGO document number LIGO-P1800072.

%% file: Cryogenics.tex
\label{s:Cryo}

\begin{figure}[t]
  \centering
  \includegraphics[width=\columnwidth]{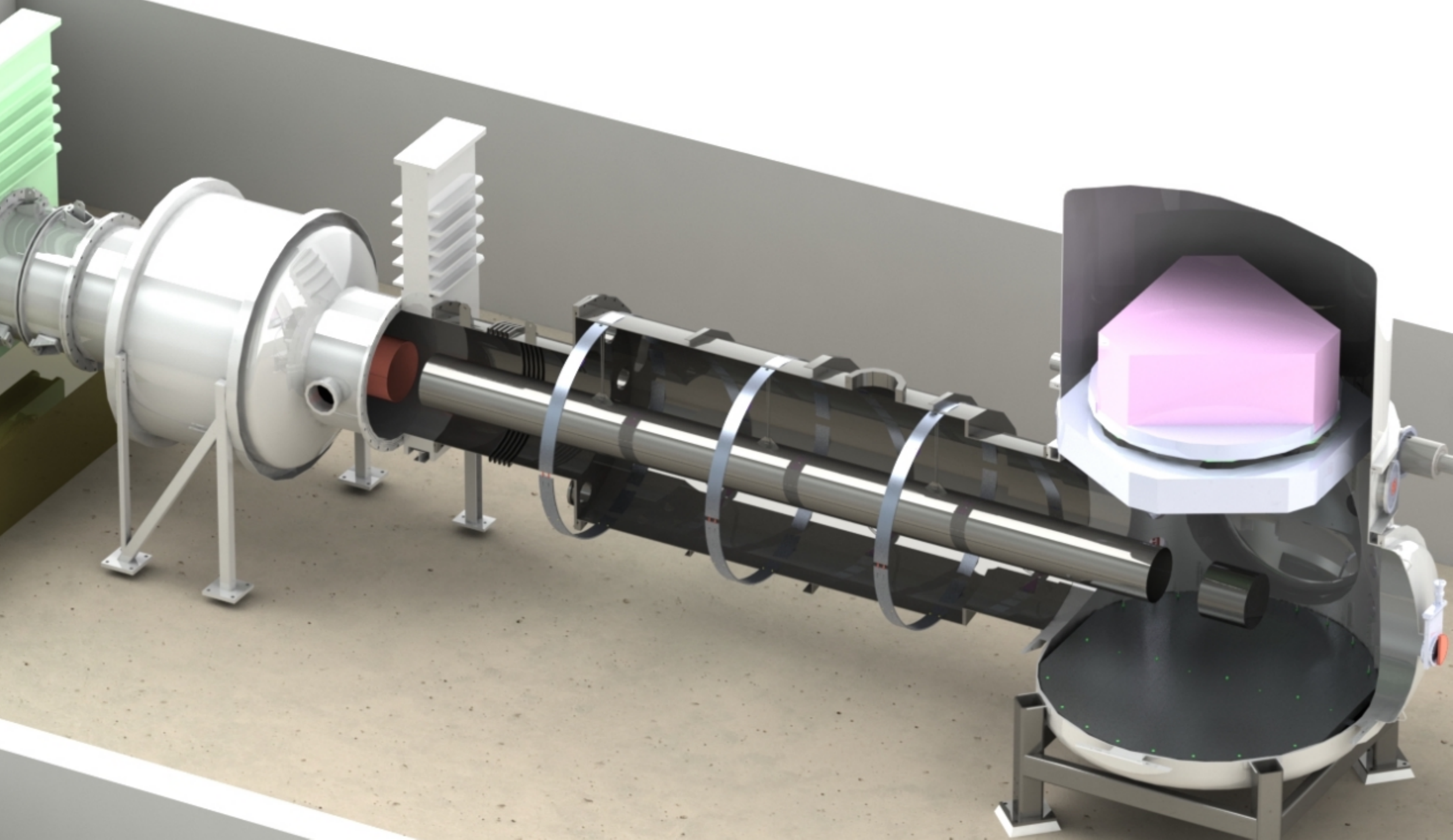}
  \caption[Drawing of end station vacuum system]
          {Rendering of end station vacuum system.
            Outer shield, mirror shield, reaction chain, and suspension cage
            structure \emph{not shown} for clarity.}
  \label{fig:endstationvac}
\end{figure}

The cryogenic design concept for \voy{} is discussed in~\cite{Shapiro2017}. The test masses are maintained at \Tcryo\,K through radiative cooling, as illustrated in Figure \ref{fig:endstationvac}. This approach was chosen to avoid complicating the test mass suspensions with the added requirements of a thermally conductive heat path.

A conductive heat path would require a mechanical link from the test mass through the suspension wires to the surrounding environment. Such a link would necessarily couple environmental vibrations into the suspension and through to the test mass. Moreover, minimizing the suspension wire cross-section is desirable to reduce the suspension thermal noise, but this would be at odds with the cryogenic design requirements (which favor thick, highly thermally conductive wires).

\subsection{Heat Loads}
The heat budget for the test mass includes several significant sources, which must be managed so as not to exceed the available radiative cooling power:
\begin{enumerate}
\item absorption of the laser beam in the high reflectivity mirror coatings
\item absorption of the laser beam in the bulk of the ITM silicon substrate
\item thermal radiation from the room temperature, 4\,km beam tube
\item thermal radiation from the vacuum chambers near the test masses
\item thermal radiation from nearby optics (reaction mass, compensation plate, arm cavity transmission monitor)
\end{enumerate}

\subsubsection{Absorption of the laser beam}
Even 1\,ppm of absorption in the high reflectivity coatings of the test masses will deliver significant heating, due to the large circulating power in the Fabry-Perot arm cavities.
Assuming a circulating arm power $P_{\mathrm{cav}} = 3\,\mathrm{MW}$, and coating absorption $\alpha_{\mathrm{C}} = 1\,\mathrm{ppm}$, the heat deposited into each test mass is 

\begin{equation}
P_{\mathrm{coating}} = P_\mathrm{cav}\,\alpha_{\mathrm{C}} = 3\,\mathrm{W}
\end{equation}

The input test masses of the arm cavities are transited by the circulating power in the power recycling cavity. Assuming a power incident on the beamsplitter $P_{\mathrm{BS}} = 3\,\mathrm{kW}$, and substrate absorption $\alpha_{\mathrm{S}} = 20\,\mathrm{ppm}/\mathrm{cm}$ in a test mass of depth $h_{\mathrm{TM}} = 55\,\mathrm{cm}$, the heat deposited into each test mass is

\begin{equation}
    P_{\mathrm{substrate}} = P_{\mathrm{BS}}\,\alpha_{\mathrm{S}}\,h_{\mathrm{TM}} = 3.3\,\mathrm{W}
\end{equation}

\subsubsection{Ambient environmental heating of the test mass}
Cold windows in the arm cavities would prevent the mirrors from being exposed to the room temperature vacuum beam tube, but are not possible to include, for several reasons. First, the Fresnel reflections from even the best AR coatings would be in excess of the acceptable arm cavity loss of 10\,ppm. Second, the beam heating of the window from the 3\,MW of circulating power would introduce a large thermal lens, which would change as the circulating power is varied. Finally, the Brownian and thermo-optic noise of a window in the arm cavity would exceed the noise in the test masses.

\begin{figure}
  \centering
  \includegraphics[width=\columnwidth]{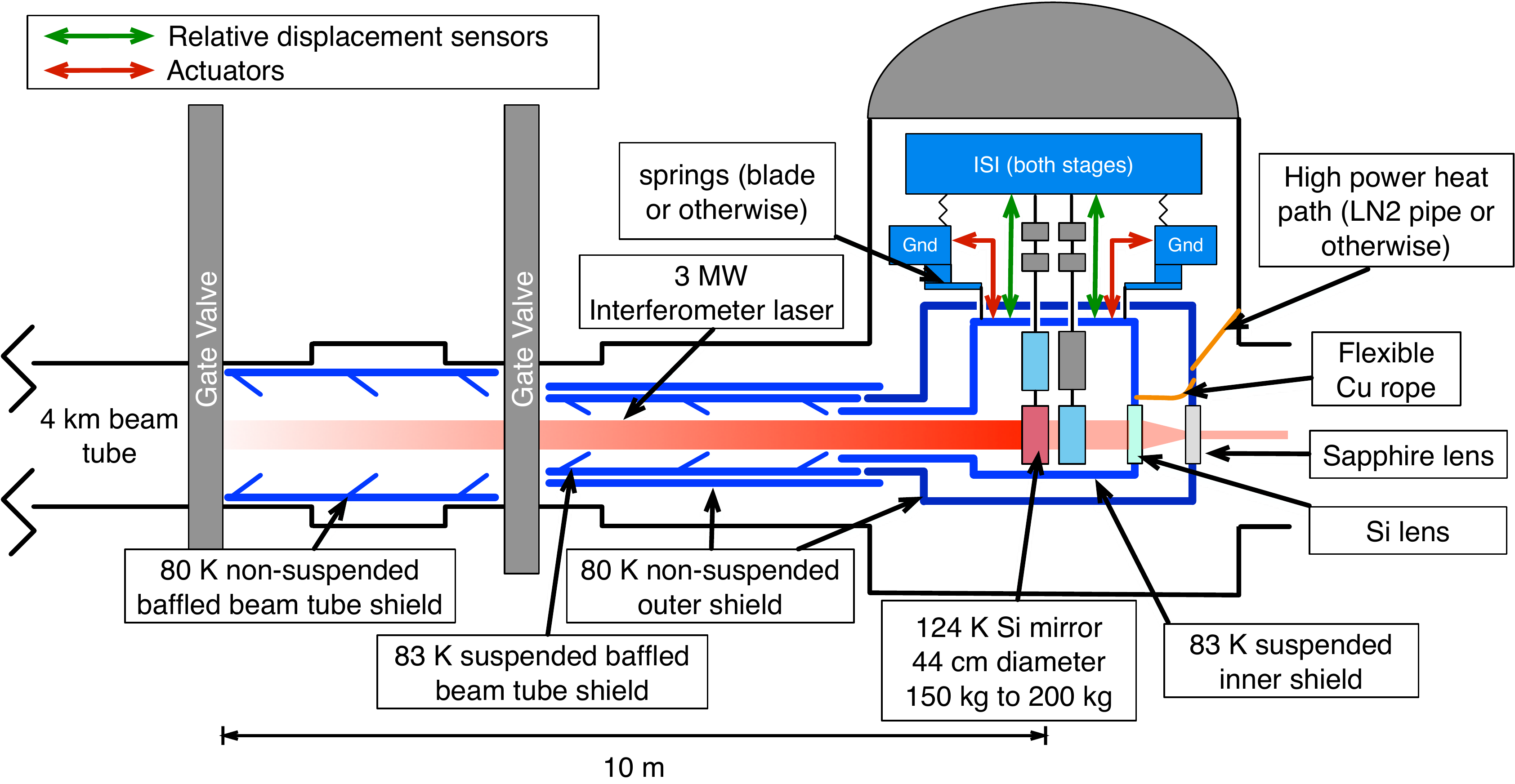}

  \caption[Suspension Cryogenics schematic with a col PUM]
  {Layout of the ETM suspension in the cold penultimate mass configuration
    with the cryogenic cooling elements. The test mass and
    reaction mass are cooled radiatively with a two layer heat
    shield system. The inner shield requires simple vibration isolation
    to mitigate scattered light noise.
    Flexible thermal straps thermally link
    the inner and outer shields to the cold head of the cryo cooler.}
  \label{fig:cryoETM_coldPUM}
\end{figure}

The radiant heating of the test mass can be largely mitigated by a cylindrical cold shield, extending out from the test mass to limit the solid angle at room temperature that the test mass `sees'. However, this shield cannot extend farther than the final gate valve separating the arm volume from the end station volume, at a distance of $\sim$\,10\,m, as illustrated in Figure \ref{fig:cryoETM_coldPUM}. The residual heating is given by the Stefan-Boltzmann law multiplied by the fraction of the full sphere subtended by the opening of the cylinder:

\begin{equation}
 P_\mathrm{beamtube} = 
\sigma\, T^4\, \pi\, r_\mathrm{TM}^2 \,\frac{\pi r_\mathrm{snout}^2}{4 \pi L_\mathrm{snout}^2}
= 6\,\mathrm{mW},
\end{equation}
assuming that the length of the shield is $L_\mathrm{snout} = 15$\,m and the radius is $r_\mathrm{snout} = 0.25$\,m.  This must be corrected for the non-black body emissivity of the HR surface.
These parameters allow the heat load from the 300\,K beam tube to be negligible. 
\begin{figure}[ht]
  \centering
  \includegraphics[width=\columnwidth]{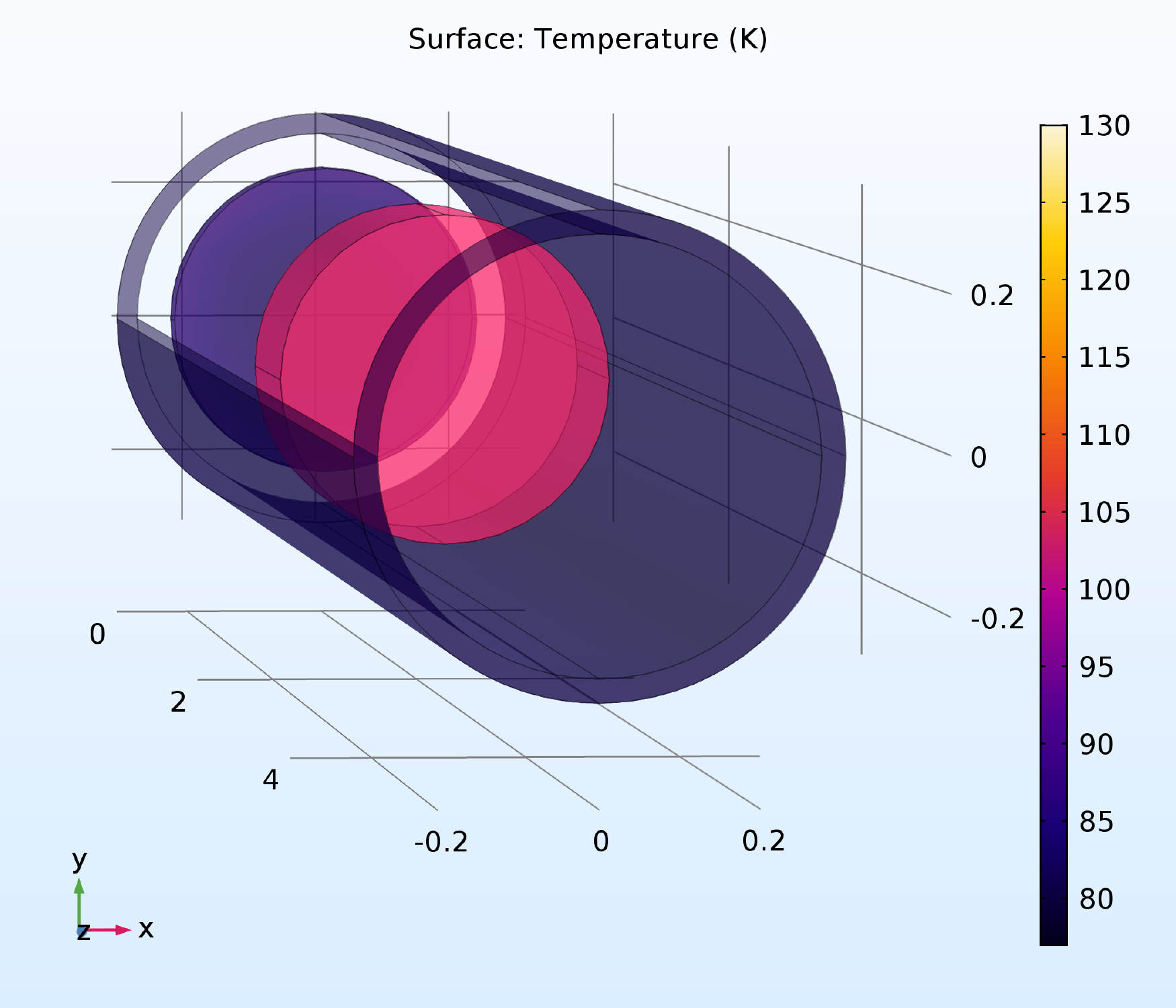}
  \caption{Cutaway view of thermal finite element model of the input test mass. The model includes heating from the main laser beam in the coating and substrate, as well as radiative heating/cooling from the surroundings.}
\label{fig:radcooling}
\end{figure}

\subsection{Radiative cooling of the test mass}
The effect of radiative cooling of the test mass into a 60\,K environment has been estimated using a finite element model (see~\Cref{fig:radcooling}).  The model presumes that the HR and AR surfaces have emissivity $\varepsilon_{\mathrm{face}} = 0.5$, and the barrel has an emissivity of $\varepsilon_{\mathrm{barrel}} = 0.9$.
At \Tcryo\,K, the test mass can radiate $\sim$10\,W.

\subsubsection{Cold Shields}
To minimize the radiative heat load from the 300\,K beam tube, the radiation shield will need to include a cylindrical piece which extends into the beam tube. The inside of the shield will require baffles, as in the KAGRA design, to reduce multiple reflection paths from the 300\,K environment~\cite{kagra_snout}.

The inside of the long shield should be coated with a high emissivity black coating to maximize the radiative coupling to the test mass.
However, there is also the consideration of the 2\,$\mu$m light scattered from the arm cavity into the shield, and then scattered back into the arm cavity.
This will be a source of amplitude and phase noise, and it is important that the high emissivity coating also has low BRDF so that scattered light noise is insignificant. Such an effect might be mitigated through the use of
a combination of specular baffling and broadband absorption.

A second shield will be used outside of these blackened inner shields to reduce the large heat load from the 300\,K environment. Both of the shields can be cooled conductively using soft thermal straps, which, in turn, are connected to Gifford-McMahon cryo-coolers outside of the vacuum system. These closed cycle cryo-coolers can cool the shields to approximately 50\,--\,\SI{60}{K}, and their vibrations can be isolated from the heat shields using simple spring mass assemblies.